\documentstyle[tighten,12pt,aps,epsf,floats]{revtex}

\newcommand{\semidir}{
\makebox[1.0em]{\small $\bigcirc$}\hspace{-1.0em}\makebox[1.0em]{s}}
\newcommand{\bfpi}{\mbox{\boldmath $\pi$}}
\newcommand{\bfspi}{\mbox{\boldmath $\scriptstyle \pi$}}

\begin{document}

\draft
\preprint{}
\title{Energy level statistics of the
two-dimensional Hubbard model\\ at low filling}
\author{Henrik Bruus and Jean-Christian Angl\`es d'Auriac}
\address{Centre de Recherches sur les Tr\`es basses Temp\'eratures,\\
CNRS, BP.\ 166, F-38042 Grenoble C\'edex 9, France}
\date{October 17, 1996}
\maketitle
\begin{abstract}
The energy level statistics of the Hubbard model for
$L\!\times\!L$ square lattices ($L=3,4,5,6$) at low filling (four
electrons) is studied numerically for a wide range of the coupling
strength. All known symmetries of the model (space,
spin and pseudospin symmetry) have been taken into account explicitly
from the beginning of the calculation by projecting into symmetry
invariant subspaces. The details of this group theoretical treatment
are presented with special attention to the nongeneric case of $L=4$,
where a particular complicated space group appears. For all the
 lattices studied, a significant amount of levels within each symmetry
invariant subspaces remains degenerated, but except for $L=4$
the ground state is nondegenerate. We explain the remaining
degeneracies, which occur only for very specific interaction
independent states, and we disregard these states in the statistical
spectral analysis. The intricate structure of the Hubbard spectra
necessitates a careful unfolding procedure, which is thoroughly
discussed. Finally, we present our results for the level spacing
distribution, the number variance $\Sigma^2$, and the spectral
rigidity $\Delta_3$, which essentially all are close to the
corresponding statistics for random matrices of the Gaussian ensemble
independent of the lattice size and the coupling strength. Even very
small coupling strengths approaching the integrable zero coupling
limit lead to the Gaussian ensemble statistics stressing the
nonperturbative nature of the Hubbard model.
\end{abstract}


\narrowtext

\section{Introduction}
\label{sec:Introduction}

The behavior of strongly correlated electronic systems remains a
central problem in contemporary condensed matter physics.  Several
years of intense studies have made it clear that the necessary
theoretical skills and tools to deal with strongly correlated fermion
systems are lacking (see e.g.\ the  recent reviews by
Dagatto\cite{Dagatto} and Lieb \cite{Lieb}). Many exotic schemes have 
been invented to accommodate a suitable theoretical framework,  but
the development of a predictive general theory does not seem to be in
sight. In this state of affairs the importance of performing numerical
calculations of the ground state properties and the energy spectrum of a
given many-body Hamiltonian has grown. Computational results can lead
to the acceptance or rejection of the proposed analytical models, and
they can guide the development of new analytical approaches. 

In the Hubbard model and related models one important parameter is the
filling $\nu$. Much work has been devoted to the high density case
near half filling, since it is believed to be relevant for
high-temperature superconductivity \cite{Dagatto,Anderson}, but also
the low filling regime is of interest, e.g.\ it plays an important
role in theoretical studies of the breakdown of Fermi liquid theory in
2D \cite{Fabrizio}. In this paper we present a numerical study of the
two-dimensional Hubbard model at low filling regime $\nu < 0.25$, a
regime where the calculation is tractable. It is natural to choose
four particles as a generic case close to the simple two-particle case. The
coupling strength is used as a perturbation parameter, and we address
the question of universality in the response of strongly correlated
electron systems to this perturbation \cite{Faas}. 

We describe an efficient method which allows for numerical
calculations of the exact energy spectrum. 
This method can relatively easily be extended to 
calculations of various Green's functions and spectral functions well
suited for the study of low lying excitations and the corresponding
coherent part of the spectral densities, a topic we will study in
forthcoming work. Here, we rather study the statistical properties of the
typical high energy excitations which is related to the incoherent
background of the typical spectral functions.  More specifically, we
study the statistical properties of the spectra within the framework
of random matrix theory (RMT). RMT was developed for the study of
neutron scattering resonances in nuclear physics in the fifties and 
sixties\cite{Mehta}, but it has since been applied to a wide range of
problems in many areas of physics \cite{Haake} (e.g.\ studies of
conduction fluctuations\cite{Stone}, microwave eigenmodes
\cite{Sridhar}, and acoustical properties of solids\cite{Ellegaard}) 
and mathematics (e.g.\ studies of the distribution of the zeros of
the Riemann zeta function\cite{Berry}). Moreover,
RMT has also been applied to various types of matrix ensembles like
Hamiltonian matrices\cite{Mehta} (as in our case), scattering matrices
\cite{Stone,Smilansky}, as well as transfer matrices\cite{HMeyer}
and Glauber matrices\cite{RMelin} of statistical mechanics models. 
Recently, RMT has been employed in the study of strongly correlated
electronic systems. Examples are studies of the 2D 
$tJ$-model\cite{Montambaux}, 2D tight-binding models \cite{Berkovits},
the 1D Bethe chain\cite{Hsu}, and the 1D Luttinger liquid\cite{Melin}.
The work presented here with emphasis on mathematical and numerical
methods is an extension of this line of research. Preliminary results
of our work have been published elsewhere\cite{Bruus}.  

There are basically two ways of applying RMT. One way is to model a
relevant matrix of the given physical system with a matrix drawn from
a suitable random matrix ensemble and subsequently calculate average
properties of the system by averaging over the random matrix ensemble
according to RMT. The other way, which we employ here, consists simply
of characterizing the spectrum of a given physical system by comparing
various statistical properties of the spectrum with the corresponding
properties calculated within one of the few universal statistical
matrix ensembles of RMT. Which of these ensembles describing properly
a physical situation depends on the symmetries of the system. The
given spectrum which one analyzes is of course deterministic, but
statistical properties are given to it by considering quantities like
for example the distribution of the energy spacings where all but one
of the spectral variables have been integrated out. This is like
pseudo-random number generators which are perfectly deterministic and
nevertheless have many properties in common with random sequences.

The paper is organized as follows. In Sec.~II we introduce the Hubbard
model and the corresponding Hilbert space. In Sec.~III we introduce the
RMT quantities used in the characterization of the spectra, and we
discuss in detail the special spectral unfolding technique
necessitated by the intricate nature of the Hubbard spectra. In
Sec.~IV the entire symmetry group consisting of space, spin and
pseudospin symmetry of the model is studied, and all the corresponding
projection operators are calculated. In Sec.~V the model is
diagonalized  numerically and we study the raw spectrum, in particular
the ground state and some unexpected remaining degeneracies higher in
the spectrum. In Sec.~VI we present the result of the spectral
statistics analysis of the model, and finally, in Sec.~VII we discuss
the results and conclude. Appendices A~-~D contain mathematical
details.

\section{The Hubbard model}
\label{sec:HubbardModel}
Throughout this paper we study 
the simple one-band $L\!\times\!L$ square lattice Hubbard model with
periodic boundary conditions containing a
nearest-neighbor hopping term -$t\hat{T}$ and an on-site interaction
term $U\hat{U}$:

\begin{equation} \label{eq:HubbardModel} 
\hat{H} = -t \hat{T} + U \hat{U} = -t \sum_{\langle
i,j \rangle,\sigma}^{L^2}
\hat{c}^{\dagger}_{j\sigma} \hat{c}_{i\sigma} + U \sum_{i}^{L^2}
\hat{n}_{i\uparrow}  \hat{n}_{i\downarrow},
\end{equation}
where $\hat{c}^{\dagger}_{i\sigma}$ and $\hat{n}_{i\sigma}$
are the creation operator and the  number operator, respectively, for an
electron on site $i$ with spin $\sigma$. No disorder is present in the
model. Below half filling the dimension $N_H$ of the Hilbert space
grows rapidly as a function of $L$ and the number $N_e$ of
electrons occupying the lattice, so we have confined ourselves to the
low filling case of only four electrons, whereas we let $L$ vary. Moreover,
without loss of generality we always work in the sector where the $z$
component $S_z$ of the total spin is zero; the other $S_z$ sectors can
be reached by use of the spin ladder operators $S_+$ and $S_-$, which
commute with the Hamiltonian. The $S_z$=0 sector is the largest of the
spin sectors, and it has $N_H = [L^2(L^2-1)/2]^2$ which for $L$ =
3, 4, 5, and 6, the lattice sizes studied here, yields 1296, 14400,
90000 and 396900, respectively. The corresponding fillings $\nu \equiv
N_e/2L^2$ are 0.22, 0.13, 0.08, and 0.06.

\begin{figure}[h]
\epsfysize=60mm \centerline{\epsfbox{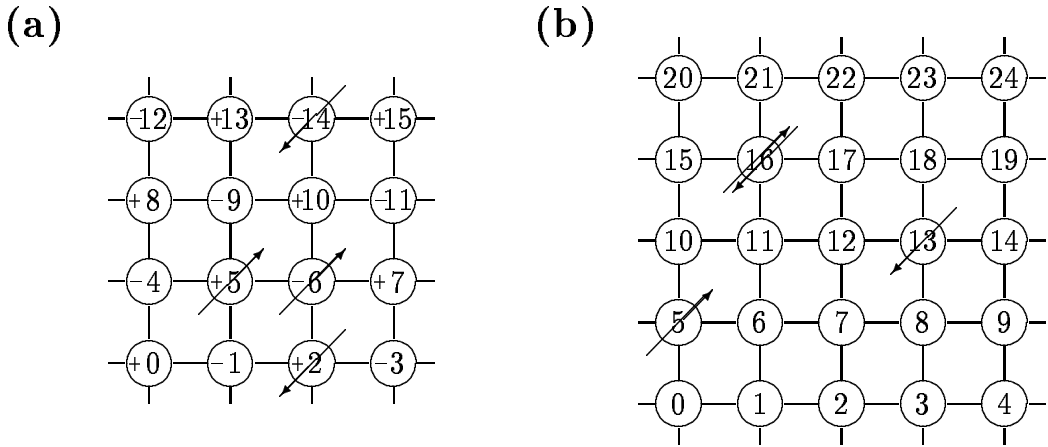}}
\caption{\label{fig:4x4and5x5}
Two four-electron states with $S_z$=0 are shown: in (a)
the zero-pair state $|5,6;2,14\rangle$ of the $4\!\times\!4$ square
 lattice, and in (b) the one-pair state $|5,16;13,16\rangle$ of the
$5\!\times\!5$ square lattice. The convention of the ket-notation is
explained in the text. The numbering of the lattice sites is the one
employed in our computer calculations. The signs of the $4\!\times\!4$
 lattice correspond to the bi-partition of the lattice, which is only
possible for even site lattices.
}
\end{figure}

In the occupation number basis we label the states as follows
\cite{Fano}, where $a$ and $b$ are two lattice sites occupied with
spin-up electrons and  $c$ and $d$ with spin-down electrons:

\begin{equation} \label{eq:ketconvention}
|a,b;c,d\rangle  \equiv 
\hat{c}^{\dagger}_{a\uparrow} \hat{c}^{\dagger}_{b\uparrow} 
\hat{c}^{\dagger}_{c\downarrow} \hat{c}^{\dagger}_{d\downarrow}
|{\rm vac}\rangle.
\end{equation}

Two explicit examples of such states as well as the lattice site
enumeration are shown in Fig.~\ref{fig:4x4and5x5}. In our work we make
sure that $a<b$ and $c<d$, and we have ordered the basis states such
that the state $|x_1,x_2;x_3,x_4\rangle$ comes before the state
$|x'_1,x'_2;x'_3,x'_4\rangle$ if $x_i<x'_i$, where $i$ is the first
position encountered where $x_i$ and $x'_i$ are different. If during a
calculation a state is encountered with $a>b$ and/or $c>d$, the
necessary exchange operations including sign changes are performed to
restore it. 

Finally we note, that for even $L$ a bi-partition of the lattice
is possible. A given lattice site $a$ can be identified by a set of
cartesian coordinates ${\bf a} = (a_1,a_2)$ simply 
counting the position in the lattice (thus e.g. site $0=(0,0)$ and
$9=(1,2)$ in Fig.~\ref{fig:4x4and5x5}a). Each site $a$ can then be
assigned with a sign $\theta_a \equiv (-1)^{a_1+a_2} =
\exp(i \bfpi \!\cdot\! {\bf a})$, where $\bfpi = (\pi,\pi)$. 

\section{Random matrix theory}
\label{sec:RMT}

Within random matrix theory (RMT) one can study several statistical
ensembles of matrices. Three important examples are the diagonal
ensembles, the Gaussian ensembles for Hermitian matrices as e.g.\
Hamiltonians, and the circular ensembles for unitary matrices as e.g.\
scattering matrices. Since a main object of this work is to
characterize the spectrum of the Hubbard model within the framework of
RMT, we are led to use the diagonal and the Gaussian ensembles of square
matrices. The first ensemble is the ensemble of diagonal matrices $D$
with statistically independent diagonal elements $D_{ii}$ drawn from
the same distribution $P(D_{ii})$. This ensemble describes situations
where the eigenvalues are essentially independent, which empirically
has been found to be the case for integrable models. One
characteristic feature of such systems is a ``soft'' spectrum with
large probabilities of having levels close together
described by the Poisson (exponential) distribution.
The three others ensembles -- denoted the Gaussian orthogonal ensemble
(GOE), Gaussian unitary ensemble (GUE), and Gaussian symplectic
ensemble (GSE) -- are defined by requirering statistical independence
of the matrix elements of a matrix $H$ and invariance of the
probability distribution $P(H)$ in matrix space under one of the three
canonical similarity transformations, the orthogonal, the unitary, and
the symplectic transformation respectively\cite{Mehta}. Which ensemble
to choose depends on the symmetry of the system. In fact, the ensembles
are universal in the sense that no details of the physical system
plays any role, only knowledge of the global symmetry is needed.
The three Gaussian ensembles are found to describe situations of very
complex or chaotic systems, and one characteristic feature of such
systems is a ``rigid'' spectrum with eigenvalue repulsion. In this
work we need only to treat GOE, which is found for systems with
preserved time-reversal symmetry.

To perform a meaningful RMT analysis one has to sort the spectrum in
symmetry sectors corresponding to the symmetry group of the
Hamiltonian since the symmetry invariant subspaces are orthogonal to
one another. Each such symmetry invariant subspace is
characterized by a specific set of quantum numbers, and the
RMT analysis is performed on sets of eigenlevels having the same
quantum numbers. In Sec.~\ref{sec:GroupTheory} we treat the
complete symmetry group of the Hubbard model and calculate the
corresponding projection operators of the symmetry invariant
subspaces. As illustrated in Sec.~\ref{sec:SpectralStat}
significant errors are introduced in the analysis if some of the
symmetries are neglected.

Another caveat in the RMT analysis of finite spectra is the notion of
mixed phase space. As a function of some external parameter a given
system can be driven from integrability (the diagonal ensemble) to
chaos (one of the Gaussian ensembles) or from one type of symmetry
(say GOE) to another (say GUE). If the spectrum of such a system is
studied in the middle of the transition the spectrum is a mixture of
two or more components each described by one of the random ensembles
\cite{Pandey}. In the thermodynamic limit usually only one
component survives, but for finite spectra this mixing calls for a further
sorting of the spectrum within each symmetry invariant subspace.
For the Hubbard model using the coupling strength $U/t$ as the
external parameter such a situation does in fact arise as the analysis
presented in Sec.~\ref{subsec:RemainDeg} shows. 

When the necessary sorting of the spectrum has been performed the
RMT analysis can begin. The first step is to unfold the spectrum.

\subsection{Unfolding the spectrum}
\label{subsec:unfolding}

Naturally, it is necessary to make some kind of transformation or
normalization of the spectrum of any given physical system to be able
to make comparisons with the universal and dimensionless results of
RMT. This operation is called the unfolding. By local rescaling of the
spectrum with the local average level spacing the unfolding transforms
the actual energies $E_i$ into dimensionless ``unfolded energies'' 
$\varepsilon_i$ with a local density of one. Thus, by unfolding
one subtracts the regular slowly varying part of the spectrum and
consider only the fluctuations. It amounts to compute from the actual
integrated density of states $N(E)$,
\begin{equation}
\label{eq:idos}
N(E)=\int_{-\infty}^E \sum_i \delta(e-E_i)\,de 
=\sum_i \theta(e-E_i),
\end{equation}
an averaged integrated density of state $\overline{N}(E)$. The unfolded
energies $\varepsilon_i$ are then given by 

\begin{equation}
\varepsilon_i = \overline{N}(E_i). 
\end{equation}
The notions of ``local density'' and ``averaged density''
are not mathematically rigorous. For some systems there exists
natural unfolding procedures. For example, it is a rigorous result
that the density of states for a $N\!\times\!N$ random matrix
approaches a semi-circular form for $N \rightarrow \infty$, thus for
any given finite random 
matrix one simply uses the limiting density of states as the average
density. In billiard systems one can make a Laurent series expansion
in $\sqrt{E}$ of $N(E)$ and obtain the Weyl law for $\overline{N}(E)$ by
truncating the series after a finite number of terms, each of which
has a physical interpretation. In our case, no such natural choices
exist for $\overline{N}(E)$. We have therefore used several 
methods to unfold the spectra. Each of these methods has a free
parameter, but there is no unique prescription of how to choose it.
The best criterion is the insensitivity of the final result to the
method employed and to reasonable variations of the free parameter.

The first method is polynomial interpolation. It can be a simple
linear interpolation or running average where $\overline{N}(E_i)$ is
found as a linear fit of $N(E)$ in an interval containing $r$ levels
on each side of the level $E_i$;  the free parameter is then the
parameter $r$. It can also be higher order polynomial interpolation,
e.g.\ in the form of interpolating between several linear
interpolations -- a method we used in our work on statistical
mechanics models \cite{HMeyer,HMeyer2}. 

The second method \cite{Haake} defines $\varepsilon_i =
\varepsilon_{i-1} + d_1/d_n$, where $d_k$ is the $k$'th smallest
spacing to $E_i$. Here $n$ becomes the free parameter. This method
works also for complex eigenvalues of non-Hermitian matrices.

The third method is Fourier broadening of the step-functions
$\theta(e-E_i)$ in Eq.~(\ref{eq:idos}). The Fourier transforms from
the energy domain to the time domain of the step-functions are found.
In the following back-transformation yielding $\overline{N}(E)$ only the
slow time components are kept. Choosing a cut-off $\tau$ beyond which
all Fourier components are set to zero yields $\theta(e-E_i) \approx
{\rm Si}[(e+E^*-E_i)\tau]/\pi - {\rm Si}[(e-E^*+E_i)\tau]/\pi$, where
${\rm Si}(x)$ is the sine integral and $E^*$ is an energy slightly
larger than the largest energy in the spectrum to be unfolded. The
free parameter is $\tau$, and by choosing $1/\tau$ to be of the order
of the mean-level spacing a good $\overline{N}(E)$ is obtained.

The fourth method is Gaussian broadening of the delta-functions
$\delta(e-E_i)$ in Eq.~(\ref{eq:idos}) leading to the following
expression for $\overline{N}(E)$:

\begin{equation}
\label{eq:idosaver}
\overline{N}(E) = \int_{-\infty}^E \sum_i \frac{1}{\sigma_i\sqrt{2 \pi}}
\exp\left[-\frac{(e-E_i)^2}{2\sigma_i^2} \right] \,de.
\end{equation}

\begin{figure}[h] 
\epsfysize=60mm \centerline{\epsfbox{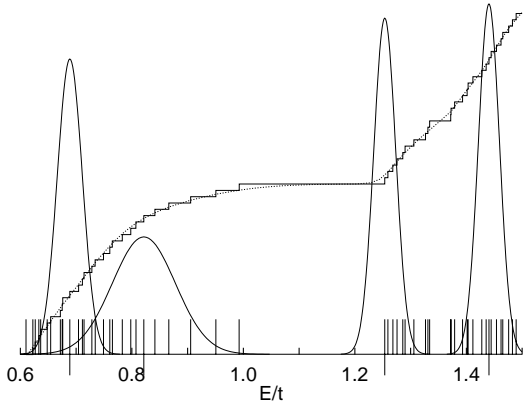}}
\caption{\label{fig:Gauss}
Unfolding of the spectrum using the Gauss broadening with variable
width choosing $\alpha=4$. Shown is a part of the Hubbard spectrum
of the invariant subspace $(R,S)=(6,0)$ around $E/t \approx 1.1$ for
$L=5$ and $U/t=1$. The positions of the levels $E_i$ are marked 
by short vertical lines. For four levels marked by long vertical lines
we show the actual broadened Gaussians. Note how the widths of the
Gaussians change as the local density of states changes, and note how
the width of the Gaussian centered near the gap ignores the states
beyond the gap. Also shown are the level staircase $N(E)$ and the
averaged integrated density of states $\protect\overline{N}(E)$ (the
smooth dotted line).
} 
\end{figure}

The standard deviation or width $\sigma_i$ of the Gaussians can be
taken as a constant for the entire spectrum; it is then the free
parameter. However, due to the appearance of many mini-bands in the
Hubbard spectrum for small values of $U/t$ each having different
densities it is desirable to let $\sigma_i$ adapt to local variations
in the spectrum. We have developed the following algorithm: take
$\alpha$ levels to each side of level $i$, determine the local average
level spacing $\Delta_i = (E_{i+\alpha} - E_{i-\alpha})/(2\alpha)$,
and set $\sigma_i = 0.608 \alpha \Delta_i$. By this assignment
90\% of the weight of the broadened peak falls in the interval $[E_i -
\alpha\Delta_i, E_i+\alpha\Delta_i]$ and $\alpha$ becomes the free
parameter. If a gap (defined as a very atypical spacing) falls within
the chosen range we only take the states of the same side of the gap
as $E_i$ into account. The procedure is illustrated in
Fig.~\ref{fig:Gauss}. We discuss how to optimize the choice of
$\alpha$ in Sec.~\ref{sec:SpectralStat}. Typically, we find $\alpha
\approx 4$ to be a good choice.

All four methods of unfolding yield essentially the same results. We
decided to use the Gaussian broadening with varying width,
Eq.~(\ref{eq:idosaver}), since it was better suited to the study of
the Hubbard spectra with its many mini-bands at small coupling
strength $U/t$ (see Sec.\ref{subsec:Spectrum}).

A final remark on the unfolded spectrum is that it is customary to
discard from the analysis the states closest to the boundary of the
spectrum or to the edges of the mini-bands. The reason is that these
levels in contrast to the levels in the bulk of the spectrum do not
interact with levels of both higher and lower energy. Hence such
levels are nongeneric. We usually discarded a few percent of the
total number of states on that account. This effect is a size effect
that is expected to be negligible in the thermodynamic limit.

\subsection{Quantities characterizing the spectrum}
\label{subsec:RMTquantities}

The simplest quantity one studies in RMT analysis is the probability
distribution $P(s)$ of unfolded energy spacings $s = \varepsilon_i -
\varepsilon_{i-1}$, where $\varepsilon_i$ and $\varepsilon_{i-1}$ are
two consecutive unfolded energies. One compares the actual $P(s)$
with the same quantity obtained for random matrices from one of the
matrix ensemble introduced in Sec.~\ref{sec:RMT}. For diagonal random
matrices $P(s)$ is the Poisson (exponential) distribution $P(s) =
\exp(-s)$. For $N\!\times\!N$ GOE matrices the distribution of
spacings is quite complicated for arbitrary $N$, however it is always
close to the exact spacing distribution of the $2\!\times\!2$ GOE
matrices known as the Wigner surmise,
\begin{equation}
\label{eq:wiglaw}
P^{\rm GOE}(s) = \frac{\pi}{2} s \exp(-\frac{\pi}{4}s^2),
\end{equation}
which therefore is used in practice.

The spacing distribution probes correlations between consecutive
states and is not sensitive to correlations of higher order. For
example an artificial spectrum constructed by adding independent
variables distributed according to Eq.~(\ref{eq:wiglaw}) will
certainly show a Wignerian spacing distribution but has clearly
nothing else in common GOE spectra. To test higher order correlations
one then looks at the two-point correlation function $Y(x)$ and
various weighted averages thereof \cite{Mehta}. For GOE in the large
$N$ limit $Y(x) = s(x)^2 + \frac{ds(x)}{dx} \int_x^\infty s(t) \,
dt$, with $s(x) = \sin(\pi x)/(\pi x)$. One average of $Y(x)$ often
studied is the number variance $\Sigma^2(\lambda)$ defined as the
variance of the number of unfolded energy
levels in intervals of length $\lambda$ around the unfolded energy
$\varepsilon_0$:

\begin{equation} \label{Sigma2def}
\Sigma^2(\lambda) = \left\langle
\left[ N_u(\varepsilon_0+\frac{\lambda}{2}) - 
       N_u(\varepsilon_0-\frac{\lambda}{2}) - \lambda \right]^2
\right\rangle_{\varepsilon_0},
\end{equation}
where $N_u(\varepsilon) \equiv \sum_i \theta(\varepsilon - \varepsilon_i)$
is the unfolded level staircase, and
where the brackets denote an averaging over $\varepsilon_0$. For the
Poissonian case $\Sigma^2(\lambda) = \lambda$, while for the GOE case
$\Sigma^2(\lambda) = \lambda - 2 \int_0^{\lambda} (\lambda - x) Y(x)
\, dx$ with a logarithmic asymptotic behavior.

Another average of the two-point correlation function is the
spectral rigidity $\Delta_3(\lambda)$ defined as the least square
deviation of the unfolded level staircase $N_u(\varepsilon)$ from the
best fitting straight line in an interval of length $\lambda$:

\begin{equation} \label{eq:Delta3def}
\Delta_3(\lambda) = \left\langle
\frac{1}\lambda \min_{(A,B)} 
\int_{\varepsilon_0-\frac{\lambda}{2}}^{\varepsilon_0+\frac{\lambda}{2}}
[N_u(\varepsilon)-A\varepsilon-B]^2d\varepsilon
\right\rangle_{\varepsilon_0} \! \! \!.
\end{equation}
For the Poissonian case the spectral rigidity is $\Delta_3(\lambda) =
\lambda/15$, while for the 
GOE case $\Delta_3(\lambda) = (\lambda - \int_{0}^{\lambda} f(x) Y(x)
\, dx)/15$, with $f(x) = (\lambda-x)^3(2\lambda^2-9\lambda x-3x^2)/
\lambda^4$ and again with a logarithmic asymptotic behavior. 

\section{Group theory and invariant subspaces}
\label{sec:GroupTheory}
The problem of diagonalizing the Hubbard Hamiltonian can be reduced
considerably by group theoretical analysis. Furthermore, as mentioned
earlier and as illustrated by an example in Sec.~\ref{subsec:OptimUnf}
it is indispensable for the RMT analysis. The symmetries are
explicitly dealt with from the beginning of the calculation by
constructing the symmetry projection operators corresponding to all
known symmetries of the model and using them to project into symmetry
invariant subspaces of the full Hilbert space \cite{Tinkham,Cornwell}.
The main object of this Section is to construct the three projection
operators ${\cal P}_R$, ${\cal P}_S$, and ${\cal P}_J$ corresponding
to the space symmetry, the spin symmetry and the pseudospin symmetry,
respectively. 

\subsection{The space symmetry group}
\label{subsec:SpaceSymm}
The first symmetry we consider is the space group $G_L$ of the $L
\!\times\! L$ square lattice with periodic boundary conditions. It
consists of all permutations $g$ of the sites such that $g(i)$ and
$g(j)$ are neighbors if and only if $i$ and $j$ are neighbors. 
In a straightforward manner an operator $\hat{g}$
in the Hilbert space can be associated 
to each element $g$ of $G_L$, $\hat{g}|a,b;c,d\rangle \equiv
|g(a),g(b);g(c),g(d)\rangle$, thus forming a group  $\hat{G}_L$ of
operators, which commutes with the Hubbard Hamiltonian $\hat{H}$.
For general values of $L$ the space group $G_L$ has been analyzed 
in detail in Ref.~\onlinecite{Fano}. Here we will restrict ourselves
to outline this analysis and to correct the particular cases of $L=2$,
which induces a simpler space group, and $L=4$, which induces a much
richer space group as briefly mentioned in Ref.~\onlinecite{HasaPoil}.

For $L\neq2,4$ the structure of $G_L$ can simply be built up
by forming direct\cite{direct} and semi-direct\cite{semidir} products,
denoted $\otimes$ and $\semidir$ respectively, of translation and
reflection subgroups. Let $T_x$ ($T_y$) be the subgroups of order $L$
of translations (isomorphic with ${\cal Z}_L$) in the $x$ ($y$)
direction, and let $r_x$, $r_y$, and $r_d$ be the reflection 
operations for the $x$ axis, the $y$ axis,  and the diagonal defined
by $r_x(x) = -x$, $r_y(y) = -y$, and $r_d(x,y) = (y,x)$,
while $e$ denotes the identity transformation. The two subgroups 
$G^x_L =  T_x \semidir \{e,r_x\}$ and 
$G^y_L =  T_y \semidir \{e,r_y\}$ of order $2L$ 
(isomorphic with $C_{Lv}={\cal Z}_L \semidir {\cal Z}_2$) 
are formed and combined into the direct
product subgroup $G^{xy}_L = G^x_L \otimes G^y_L$. 
Finally $G_L$ is formed by the semi-direct product $G_L = G^{xy}_L
\semidir \{e,r_d\}$. One can say that $G_L$ is generated by the elements
$\{t_x,r_x,t_y,r_y,r_d\}$ although this is not the smallest possible set
of generators. The order $N_L$ of $G_L = (C_{Lv} \otimes C_{L v} )
\semidir {\cal Z}_2$ is seen to be $N_L = 8L^2$.  

For $L=2$ we find $G_2 = (\{e,r_x\} \otimes \{e,r_y\})
\semidir \{e,r_d\} = C_{4v}$ with $N_2 = 8$. This result differs from
that of the general case since $t_x = r_x$ and $t_y = r_y$.

\begin{figure}[h]
\epsfysize=60mm \centerline{\epsfbox{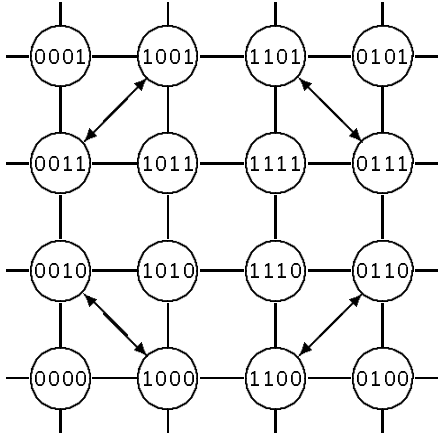}}
\caption{\label{fig:hypercube4D}
The mapping of the $4\!\times\!4$ square lattice with periodic
boundary conditions onto the unit hypercube in 4D. The four
double arrows indicate the neighbor preserving transformation $r_h$.
This transformation cannot be expressed by the ordinary translations
and reflections in 2D, however, it can be interpreted as a
hyperplane reflection in 4D around the $xz$ plane.
}
\end{figure}

For the case $L=4$ a richer group appears because the $4\!\times\!4$
 lattice with periodic boundary conditions is isomorphic with the group
of transformations of the four-dimensional unit hypercube. This
isomorphism is easily seen by changing the decimal enumeration of
Fig.~\ref{fig:4x4and5x5}a into a binary enumeration such that the binary
numbers of neighboring sites differs by only one bit as shown in
Fig.~\ref{fig:hypercube4D}. The can immediately be interpreted as the
coordinates of the corners of the hypercube. Thus the group $G_4$ of
neighbor preserving transformations are given by combining bit
inversion operations $0\!\leftrightarrow\!1$ with permutations of the
four bits. In Appendix~\ref{sec:appendixA} we show that $G_4$ is
isomorphic with the group $({\cal Z}_2 \otimes {\cal Z}_2 \otimes
{\cal Z}_2 \otimes {\cal Z}_2) \semidir S_4$ corresponding to the
semi-direct product of all combinations of bit inversions at the four
positions with the permutation group $S_4$ of the four positions. One
neighbor-preserving transformation of the lattice which is not given
by products of the elements $\{t_x,r_x,t_y,r_y,r_d\}$ is the hyperplane 
reflection $r_h$ (see Fig.~\ref{fig:hypercube4D}) obtained in 4D by a
reflection in the $xz$-plane while keeping $y$ and $w$ fixed.
$G_4$ can be generated by replacing $r_d$ above with $r_h$ (we note
that $r_d = t_x^{-1} t_y^{-1} r_h t_yt_x r_h$), and a set of
generators is  $\{t_x,r_x,t_y,r_y,r_h\}$ though in fact as few as two
elements can be found that generates $G_4$. \cite{TwoGen} Instead of
128 elements found by the formula $N_L = 8L^2$, we find $N_4 =
2^4\!\times\! 4! = 384$. We refer the reader to
Appendix~\ref{sec:appendixA} and Appendix~\ref{sec:appendixB} for
further details concerning $G_4$ and $G_L$, respectively.

The complete table of characters $\chi^{(R')}_R(g)$ of $G_4$ is
given in Table~\ref{tab:Chi4} in Appendix~\ref{sec:appendixA}, while
those of $G_3$, $G_5$, and $G_6$ are found by combining
Table~\ref{tab:Irrep356} in Appendix~\ref{sec:appendixB} with the
method of Ref.\ \onlinecite{Fano}. Armed with the characters tables
the character projection operator $\hat{\cal P}_R^{(R')}$ (where
subscript $R$ is used to distinguish from $S$ and $J$) can be
constructed for each $l_{R'}$-dimensional irreducible representation
$R'$:

\begin{equation} \label{eq:Pchi}
\hat{\cal P}_R^{(R')} \equiv \frac{l_{R'}}{N_L} \sum_{g\!\in\!G_L}
\chi_R^{(R')*}(g) \hat{g}.
\end{equation}
Using $\hat{\cal P}_R^{(R')}$ for a given representation together with
the Van Vleck basis-function generating
algorithm\cite{Tinkham,Cornwell} in the actual or a closely related
Hilbert space it is straightforward to construct an orthonormal
$l_R$-dimensional basis $\{\phi^{(R)}_1, \phi^{(R)}_2, \ldots
,\phi^{(R)}_{l_R}\}$ and calculate an explicit irreducible
representation matrix $\Gamma^{(R)}_{ij}(g) \equiv
\langle\phi^{(R)}_i|\hat{g}|\phi^{(R)}_j\rangle$. Finally the row
projection operator $\hat{\cal P}^{(R')}_{Rk}(g)$ can be constructed:

\begin{equation} \label{eq:Prow}
\hat{\cal P}^{(R')}_{Rk} \equiv \frac{l_{R'}}{N_L} \sum_{g\!\in\!G_L}
\Gamma^{(R')*}_{kk}(g) \hat{g}.
\end{equation}
Both projection operators $\hat{\cal P}_R^{(R')}$ and $\hat{\cal
P}^{(R')}_{Rk}$ will be employed in the diagonalization of the Hubbard
Hamiltonian.

\subsection{The SU(2) spin symmetry}
\label{subsec:SpinSymm}
It is easily verified that the Hubbard Hamiltonian
Eq.~(\ref{eq:HubbardModel}) commutes with the $z$ component of the
total spin operator $\hat{S}_z = \sum_{i=1}^{N_e} \hat{S}_z^{(i)}$,
as well as with the corresponding  raising and lowering operators
$\hat{S}_+$ and $\hat{S}_-$. The model therefore possesses a SU(2)
spin symmetry. We can therefore restrict our diagonalization to the
$S_z$=0 sector, since the other sectors can be reached using the
spin raising and lowering operators. The spin operators also commute
with any space symmetry operator $\hat{g}$, so the two symmetry groups
form a direct product group. The combination of two spin-up and two
spin-down electrons under the constraint $S_z=0$ leads to a total spin
quantum number $S'$ = 0, 1 or 2. The explicit 
construction in Appendix~\ref{sec:appendixC} with $\hat{O} = \hat{\bf
S}^2$ yields the following three spin projection operators $\hat{\cal
P}^{(S')}_S$:

\begin{eqnarray}
\label{eq:PS0}
\hat{\cal P}^{(0)}_S|a,b;c,d\rangle &\!=\!& 
\frac{1}{6}(
     2|a,b;c,d\rangle
\!+\!2|c,d;a,b\rangle
\!+\! |a,c;b,d\rangle
\!+\! |b,d;a,c\rangle
\!-\! |a,d;b,c\rangle
\!-\! |b,c;a,d\rangle)\\
\label{eq:PS1}
\hat{\cal P}^{(1)}_S|a,b;c,d\rangle &\!=\!& 
\frac{1}{6}(
     3|a,b;c,d\rangle
\!-\!3|c,d;a,b\rangle)\\
\label{eq:PS2}
\hat{\cal P}^{(2)}_S|a,b;c,d\rangle &\!=\!& 
 \frac{1}{6}(
     1|a,b;c,d\rangle
\!+\!1|c,d;a,b\rangle
\!-\! |a,c;b,d\rangle
\!-\! |b,d;a,c\rangle
\!+\! |a,d;b,c\rangle
\!+\! |b,c;a,d\rangle).
\end{eqnarray}
These expressions are generally valid, however, one should note that
the Pauli exclusion principle reduces the number of terms when
electron pairs are present. 
For example if $a\!=\!c$ and $b\!=\!d$ one finds ${\cal
P}_S^{(S')}|a,b;c,d\rangle = \delta_{S',0} |a,b;c,d\rangle$. 
Finally we note, that the spin projectors of the state
$|a,b;c,d\rangle$ involve all six states generated by permutations of
the  site indices. In Eqs.~(\ref{eq:PS0})-(\ref{eq:PS2}) only three
orthogonal states appear. The remaining three orthogonal states are
the two $S$=1 states, $|a,c;b,d\rangle \!-\! |b,d;a,c\rangle$ and 
$|a,d;b,c\rangle \!-\! |b,c;a,d\rangle$, and the one $S$=0 state
$|a,c;b,d\rangle\!+\! |b,d;a,c\rangle \!+\! |a,d;b,c\rangle
\!+\! |b,c;a,d\rangle$.

\subsection{The SU(2) pseudospin symmetry}
\label{subsec:PseudospinSymm}
The pseudospin symmetry of the Hubbard model has been known for at
least a quarter of century\cite{Heilmann}, but recently it was
rediscovered and put in the more generalized context of the
$\eta$-pairing mechanism\cite{CNYang,YangZhang}. As discussed in
Sec.~\ref{sec:HubbardModel} even $L$ lattices can be bi-parted
using the site sign $\theta_i$ as an index, and three operators
$\hat{J}_-$, $\hat{J}_+$, and $\hat{J}_z$ can be defined:

\pagebreak

\begin{equation} \label{eq:Jdef}
\hat{J}_- =\sum_i\theta_i 
\hat{c}_{i\uparrow} \hat{c}_{i\downarrow} 
\hspace*{10mm}
\hat{J}_+ =\sum_i\theta_i
\hat{c}_{i\downarrow}^{\dagger} \hat{c}_{i\uparrow}^{\dagger} 
\hspace*{10mm}
\hat{J}_z = -\frac{1}{2}
\left( L^2  - \sum_{i\sigma} \hat{n}_{i\sigma} \right).  
\end{equation}
It is seen that $\hat{J}_+$ creates a pair of electrons with phase
$\theta_i$ on empty sites $i$. 
These three operators form the same algebra as $\hat{S}_-$,
$\hat{S}_+$, and $\hat{S}_z$, hence the name pseudospin. 
Furthermore, the $\hat{J}$-operators commutes with both the space
symmetry operators and the spin operators; and for the symmetrized
Hubbard model $\hat{H}'$ \cite{SymHubModel}, trivially related to our
model $\hat{H}$, they even commute with the Hamiltonian $\hat{H}'$.
Hence the (symmetrized) model possesses an extra SU(2)
symmetry characterized by the quantum numbers $J'$ and $J_z$ analogous
to $S'$ and $S_z$ for the spin \cite{SvsJ}. A detailed analysis
shows that the combination of the spin and the pseudospin symmetry
yields [SU(2)$\otimes$SU(2)]/${\cal Z}_2$ = SO(4) rather than the full
SU(2)$\otimes$SU(2) \cite{YangZhang}. The space symmetry  
group is completely independent of the spin and pseudospin symmetries.
Thus the full symmetry group for $L$ even is ${\cal G} = G_L\otimes{\rm
SO(4)}$ while for $L$ odd it is ${\cal G} = G_L\otimes{\rm SU(2)}$. In
our case the symmetry is lowered in a trivial way from spherical to 
cylindrical symmetry in pseudospin space since $[\hat{U},
\hat{J}_{\pm}] = \pm \hat{J}_{\pm}$, but still we have
$[\hat{H},\hat{J}^2] = [\hat{H},\hat{J}_z] = 0$ and $J'$ and $J_z$
both remain good quantum numbers.

For a given lattice with even $L$ containing a fixed number $N_e$ of
electrons all states have $J_z = (N_e-L^2)/2$ (e.g.\ $-6$ for $L$=4
and $-16$ for $L$=6). Defining $J_0 \equiv |J_z|$, the size $J$ of the
pseudospin in this situation takes the values $J_0, J_0+1, \ldots ,
J_0+N_e/2$. The projection operators $\hat{\cal P}^{(J')}_J$ are found
using the construction of Appendix~\ref{sec:appendixC} with $\hat{O} =
\hat{J}^2 = \hat{J}_+\hat{J}_- + \hat{J}_z^2 - \hat{J}_z$. Due to the 
explicit reference to pairs in the definition of the pseudospin
operators, it has proven useful to introduce a special notation for
one-pair and two-pair basis states as follows:

\begin{equation}
\label{eq:PairStates}
\begin{array}{rcrcr}
|P;b,d\rangle & \equiv & 
\theta_P c^{\dagger}_{P\downarrow}c^{\dagger}_{P\uparrow} 
c^{\dagger}_{b\uparrow}c^{\dagger}_{d\downarrow}|{\rm vac}\rangle
& = & \theta_P|P,b;P,d\rangle\\
|P;Q\rangle & \equiv & 
\theta_P \theta_Q c^{\dagger}_{P\downarrow}c^{\dagger}_{P\uparrow} 
c^{\dagger}_{Q\downarrow}c^{\dagger}_{Q\uparrow}|{\rm vac}\rangle
& = & -\theta_P \theta_Q |P,Q;P,Q\rangle
\end{array}
\end{equation}
For zero-pair states where $a$, $b$, $c$, and $d$ are all different,
only $\hat{\cal P}^{(J'=J_0)}_J$ is nonzero,

\begin{equation}\label{eq:PJ00}
\hat{\cal P}^{(J_0)}_J|a,b;c,d\rangle = |a,b;c,d\rangle,
\end{equation}
for one-pair states where $P$,$b$, and $d$ are all different, two
projection operators are nonzero,

\begin{eqnarray} 
\label{eq:PJ10} {\displaystyle
\hat{\cal P}^{(J_0)}_J|P;b,d\rangle} & = & 
{\displaystyle
\frac{2J_0+1}{2(J_0+1)}|P;b,d\rangle +
\frac{-1}{2(J_0+1)} \sum_{Q\neq P,b,d} |Q;b,d\rangle},\\[3mm]
\label{eq:PJ11} {\displaystyle
\hat{\cal P}^{(J_0+1)}_J|P;b,d\rangle} & = & 
{\displaystyle
\frac{1}{2(J_0+1)}|P;b,d\rangle +
\frac{1}{2(J_0+1)} \sum_{Q\neq P,b,d} |Q;b,d\rangle},
\end{eqnarray}
while for two-pair states where $P \neq Q$, all three projection
operators are nonzero: 

\begin{eqnarray} 
{\displaystyle
\hat{\cal P}^{(J_0)}_J|P;Q\rangle} & = & 
{\displaystyle
\frac{2J_0+1}{2J_0+3}|P;Q\rangle +
\frac{-(2J_0+1)}{2(J_0+1)(2J_0+3)} \sum_{R\neq P,Q} 
\{|R;Q\rangle+|R;P\rangle\}} \nonumber \\[2mm]
\label{eq:PJ20} && \hspace*{6.5em} {\displaystyle +
\frac{1}{2(J_0+1)(2J_0+3)} 
\sum_{R\neq P,Q}\sum_{S\neq P,Q,R} |R;S\rangle},\\[3mm]
{\displaystyle
\hat{\cal P}^{(J_0+1)}_J|P;Q\rangle} & = & 
{\displaystyle
\frac{1}{J_0+2}|P;Q\rangle +
\frac{J_0}{2(J_0+1)(J_0+2)} \sum_{R\neq P,Q} 
\{|R;Q\rangle+|R;P\rangle\}} \nonumber\\[2mm]
\label{eq:PJ21} && \hspace*{6.0em}  {\displaystyle +
\frac{-1}{2(J_0+1)(J_0+2)} 
\sum_{R\neq P,Q}\sum_{S\neq P,Q,R} |R;S\rangle},\\[3mm]
{\displaystyle
\hat{\cal P}^{(J_0+2)}_J|P;Q\rangle} & = & 
{\displaystyle
\frac{1}{(J_0+2)(2J_0+3)}|P;Q\rangle +
\frac{1}{(J_0+2)(2J_0+3)} \sum_{R\neq P,Q} 
\{|R;Q\rangle+|R;P\rangle\}} \nonumber \\[2mm]
\label{eq:PJ22} && \hspace*{10.5em} {\displaystyle +
\frac{1}{2(J_0+2)(2J_0+3)} 
\sum_{R\neq P,Q}\sum_{S\neq P,Q,R} |R;S\rangle}.
\end{eqnarray}
At this stage all the group theoretical ingredients are ready for the
diagonalization of the Hubbard Hamiltonian.

\subsection{The symmetry invariant subspaces}
\label{subsec:Subspaces}
Using the projection operators the $N_H$-dimensional Hilbert space can
be broken down into smaller symmetry invariant subspaces. Let $\cal G$
be the group containing the symmetry operations $\gamma$, each being a
product of a space symmetry transformation, a spin rotation and  (for
$L$ even) a pseudospin rotation, $\gamma = g \otimes g_S \otimes g_J$.
${\cal G}$ thus consists of one finite 
group  and one or two compact Lie groups. Let $\rho$ be a multi-index
describing an irreducible representation $\Gamma^{(\rho)}$ of ${\cal G}$. 
For even (odd) $L$ we have $\rho = (R',S',J')$ ($\rho = (R',S')$). The
``celebrated'' theorem\cite{Tinkham,Cornwell} states how many times
$a_{\rho}$ each row of the irreducible representation
$\Gamma^{(\rho)}$ with character $\chi^{(\rho)}$ appears in any given
not necessarily irreducible representation $\Gamma$ with character
$\chi$:

\begin{equation} \label{eq:CelebratedTheorem}
a_{\rho} =  \int_{\cal G} d\gamma \chi^{(\rho)*}(\gamma) \chi(\gamma) 
= \frac{1}{N_L}\sum_{G_L} \int_{\rm SU(2)} dg_S \int_{\rm SU(2)} dg_J
\chi^{(\rho)*}(\gamma) \chi(\gamma) 
\end{equation}
We now chose $\Gamma$ to be the following $N_H$-dimensional reducible
representation: 

\begin{equation} \label{eq:ReducibleRep}
\Gamma_{ij}(\gamma) \equiv \langle i | \hat{\gamma} |j \rangle
\hspace{10mm} 
\chi(\gamma) = \sum_{i=1}^{N_H}  \langle i | \hat{\gamma} |i \rangle.
\end{equation}
Inserting into Eq.~(\ref{eq:CelebratedTheorem}) these $\chi(\gamma)$'s
which are directly linked to the Hilbert space  yields:

\begin{equation} \label{eq:a_rho}
a_{\rho} =  \int_{\cal G} d\gamma  \chi^{(\rho)*}(\gamma) 
\sum_{i=1}^{N_H}  \langle i | \hat{\gamma} |i \rangle
= \frac{1}{l_R} \sum_{i=1}^{N_H} \langle i | 
\hat{\cal P}^{(\rho)} | i \rangle
= \frac{1}{l_R} \sum_{i=1}^{N_H} \langle i | 
\hat{\cal P}_R^{(R')} \otimes \hat{\cal P}_S^{(S')}  
\otimes \hat{\cal P}_J^{(J')}
| i \rangle,
\end{equation}
which is obtained by using $\chi(\gamma) = 
\chi(g\!\otimes\!g_S\!\otimes\!g_J) = \chi(g) \chi(g_S) \chi(g_J$)
and the definitions of the character projection operators. Note that
only the space group projection needs a normalizing factor $1/l_R$.
The spin and pseudospin are both restricted to take only one value of
their respective $z$ component; hence their normalizing factor is 1.

The initial size of the Hubbard Hamiltonian to be diagonalized is
$[L(L-1)/2]^2$. However, only states transforming according to the
same row of the same irreducible representation $\rho$ can have a
nonzero matrix element. Hence the Hamiltonian matrix breaks up in
blocks, one per representation, and within each representation a
further division into $l_R$ equivalent blocks occurs. The relationship
between $N_H$, $a_{(R',S',J')}$, and $l_{R'}$ is $N_H =
\sum_{(R,S,J)} l_{R'} a_{(R',S',J')}$. This reduction is considerable as
shown in Table~\ref{tab:SizeReduction}.

\begin{table}[htb]
\begin{tabular}{|c||cccc|}
& $L$ = 3 & $L$ = 4 & $L$ = 5 & $L$ = 6 \\ \hline \hline
$N_H$ & 1296 & 14400 & 90000 & 396900 \\
max$_\rho$\{$a_{\rho}$\} & 38 & 146 & 1794 & 5490 \\ 
$\sum_{\rho} a_{\rho}^3/N_H^3$ 
& $14.1\!\times\!10^{-5}$
& $0.7\!\times\!10^{-5}$
& $2.3\!\times\!10^{-5}$
& $1.0\!\times\!10^{-5}$ \\
\end{tabular}
\caption{\label{tab:SizeReduction}
For  $L=3,4,5$, and 6 are shown the dimension 
$N_H$ of the total unreduced Hilbert space and the dimension
max\{$a_{\rho}$\} of the largest symmetry invariant subspace.
Since matrix diagonalization is a $n^3$-operation, the number
$\sum_{\rho} a_{\rho}^3/N_H^3$ provides an estimate of the relative
reduction in computer time by projecting into the invariant subspaces.}
\end{table}

\section{Numerical diagonalization of the Hubbard model}
\label{sec:NumericalDiag}
The numerical calculation of the exact spectra of the
Hubbard model begins by determining the block size $a_{\rho}$ for all
the irreducible representations $\rho$. Then for a given $\rho = 
(R',S',J')$, where $J'$ is disregarded in case of odd $L$, an
$a_{\rho}$-dimensional orthonormal basis is found by using the
projection operator  

\begin{equation} \label{eq:FinalProjection}
\hat{\cal P}_0^{(\rho)} \equiv \
\hat{\cal P}_{R0}^{(R')} \otimes
\hat{\cal P}_S^{(S')}  \otimes
\hat{\cal P}_J^{(J')},
\end{equation}
which projects onto the first row of the space group representation
$R'$ with the spin index $S'$ and pseudospin index $J'$. The
projection operator $\hat{\cal P}_0^{(\rho)}$ is applied on one basis
state after another while performing an ongoing Gram-Schmidt
orthonormalization procedure. This yields new basis states
$|w_i\rangle$, and when $a_{\rho}$ such states are found, the
procedure is terminated.

Next step is to calculate the kinetic energy and potential energy
matrix elements  $\langle w_i | \hat{T} | w_j \rangle$ and $\langle w_i
| \hat{U} | w_j \rangle$. These matrix elements are stored in the
computer and thereafter it is a simple matter to pick any value of
$U/t$ and diagonalize the Hubbard Hamiltonian, $\hat{H} = -t\hat{T} +
U\hat{U}$,  using standard diagonalization routines. The calculation
of the symmetry invariant matrix blocks of $\hat{T}$ and $\hat{U}$
takes time of the order of one diagonalization.

\subsection{The spectrum and first order perturbation theory of the
ground state}
\label{subsec:Spectrum}
\begin{figure}[htb]
\epsfxsize=\textwidth \centerline{\epsfbox{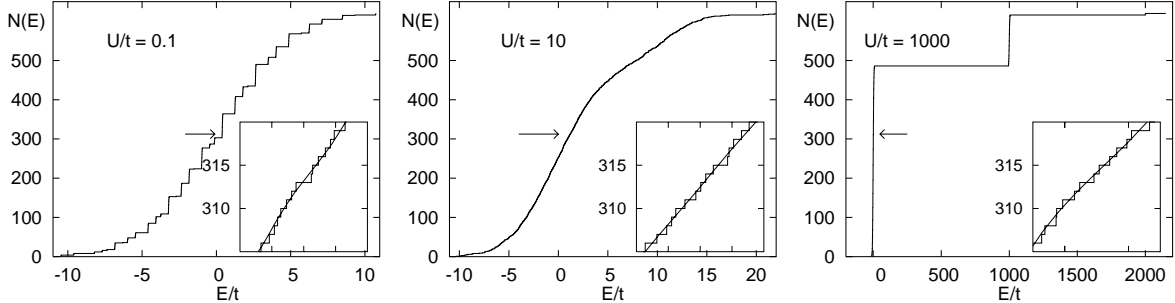}}
\caption{\label{fig:ThreeSpectra}
The integrated density of states $N(E)$ of the Hubbard model with 4
electrons on the $5\!\times\!5$ lattice for the irreducible
representation $(R,S) = (10,0)$ with $U/t = 0.1,10,1000$. For
$U/t=0.1$ the mini bands are still clearly visible, $N(E)$ is
essentially featureless for $U/t=10$, while for $U/t=1000$ the three 
Hubbard bands at 0, 1000 and 2000 appears. The inserts are
magnifications of $N(E)$ where the arrows point. The smooth
$\overline{N}(E)$ is added.
}
\end{figure}

In this section we discuss some features of the raw spectrum of the
Hubbard model. Only later in Sec.~\ref{sec:SpectralStat} we are going
to unfold the spectrum and look for universal features.
As a function of the coupling strength $U/t$ the spectrum clearly
falls in three classes. In the weak coupling limit ($U/t \ll W/t$),
to be studied in more detail below, the spectrum acquires a band width
$W \approx 32 t$. It consists of a number $M$ of well separated mini
bands reminiscent of the huge degeneracy at zero coupling due to size
quantization. For intermediate coupling strengths ($U/t \approx W/t$)
the spectrum becomes rather featureless. No apparent gaps or bands
emerge. In the strong coupling limit ($U/t \gg W/t$) the
spectrum splits up in three well separated bands centered around the
energies $E/t = 0,U/t,2U/t$. These are the Hubbard bands
corresponding to states containing zero, one or two pairs of
electrons. In Fig.~\ref{fig:ThreeSpectra} are shown one typical
spectrum from  each of the three coupling strength regimes.

We now focus on the weak coupling limit where it is natural to make
Fourier transforms from real space to momentum space,
$\hat{c}^{\dagger}_{{\bf k}\sigma} \equiv \frac{1}{L} 
\sum_{j} \exp(i{\bf k}\!\cdot\!{\bf r}_j)
\hat{c}^{\dagger}_{j\sigma}$, where ${\bf k} = (k^x,k^y)$, with
$k^{\zeta} = 2 \pi n/L$ and $n=0,1,\ldots, L-1$. 
The eigenstates of the kinetic energy operator $-t\hat{T}$ are

\begin{eqnarray} 
\nonumber 
{\displaystyle
|{\bf k}_0,{\bf k}_1;{\bf k}_2,{\bf k}_3\rangle}
& = &
{\displaystyle
\hat{c}^{\dagger}_{{\bf k}_1\uparrow}
\hat{c}^{\dagger}_{{\bf k}_2\uparrow}
\hat{c}^{\dagger}_{{\bf k}_3\downarrow}
\hat{c}^{\dagger}_{{\bf k}_3\downarrow} |{\rm vac}\rangle} \\
\label{eq:ZeroEigen}
{\displaystyle
-t\hat{T}|{\bf k}_0,{\bf k}_1;{\bf k}_2,{\bf k}_3\rangle} 
& = &
{\displaystyle
-2t \sum_{n=0}^{3} [\cos(k^x_n) + \cos(k^y_n)]
|{\bf k}_0,{\bf k}_1;{\bf k}_2,{\bf k}_3\rangle.}
\end{eqnarray}
The Pauli principle prevents 
${\bf k}_0$=${\bf k}_1$ and ${\bf k}_2$=${\bf k}_3$, and the ground state
energy $E^{(0)}_L$ becomes 

\begin{equation} \label{Erg0}
E^{(0)}_L = -12t-4t\cos(2\pi\!/\!L). 
\end{equation}
For large lattices this tends toward $-16t$ and results in a band
width $W = 32t$. For $L$ = 3, 4, 5, and 6, respectively, the actual
band widths $W/t$ are 20.0, 24.0, 26.2, and 28.0 while the number $M$
of mini bands for $U=0$ are 7, 13, 42, and  29.

The limitation of perturbation theory is demonstrated by a first order
degenerate perturbation calculation for the ground state. Let $p_L$ be
the one-dimensional momentum component $p_L = 2\pi/L$ and construct
the five two-dimensional momentum vectors ${\bf 0} = (0,0)$, 
${\bf q}^0 = (p_L,0)$, 
${\bf q}^1 = (0,p_L)$, ${\bf q}^2 = (-p_L,0)$, and ${\bf q}^3 =
(0,-p_L)$. The 16 states $|\mu,\lambda\rangle$, defined as

\begin{equation} \label{eq:MuLambda}
|\mu,\lambda\rangle = |{\bf 0}_{\uparrow}, {\bf q}^{\mu}_{\uparrow};
{\bf 0}_{\downarrow},{\bf q}^{\lambda}_{\downarrow}\rangle, 
\qquad \mu,\lambda = 0,1,2,3,
\end{equation}
form the ground state multiplet separated from the next multiplet by
an energy gap $\Delta E_L = 2t[1-\cos(p_L)]$.
In momentum space the interaction operator $U\hat{U}$ takes the form

\begin{equation} \label{eq:uUmomentum}
U\hat{U} = \frac{U}{L^2} 
\sum_{{\bf k}_1,{\bf k}_2,{\bf q}}
\hat{c}^{\dagger}_{{\bf k}_1+{\bf q}\uparrow}  \hat{c}_{{\bf k}_1\uparrow}
\hat{c}^{\dagger}_{{\bf k}_2-{\bf q}\downarrow}\hat{c}_{{\bf k}_2\downarrow},
\end{equation}
and after some simple algebra the matrix elements of the perturbation
are found to be

\begin{equation} \label{eq:UpertMatrix}
\langle m,l | U\hat{U} | \mu, \lambda \rangle = \frac{U}{L^2} \left( 3
\delta_{m,\mu} \delta_{l,\lambda} + \delta_{m+l,\mu+\lambda} \right).
\end{equation}
The eigenvalues can be found analytically, and we end up with the
following expression for the perturbed levels $E^{(1)}_{L,\beta}(U)$
with degeneracies $d$:

\begin{equation} \label{eq:PertErg}
E^{(1)}_{L,\beta}(U) = E^{(0)}_L + \beta \frac{U}{L^2}, \qquad 
\beta = 3_{(d=7)}, 4_{(d=4)}, 5_{(d=4)}, 7_{(d=1)}.
\end{equation}
Thus in first order perturbation theory the ground state degeneracy is
partly lifted leaving a 7-fold degenerated ground state for $U>0$.

\begin{figure}[h] 
\epsfxsize=85mm \centerline{\epsfbox{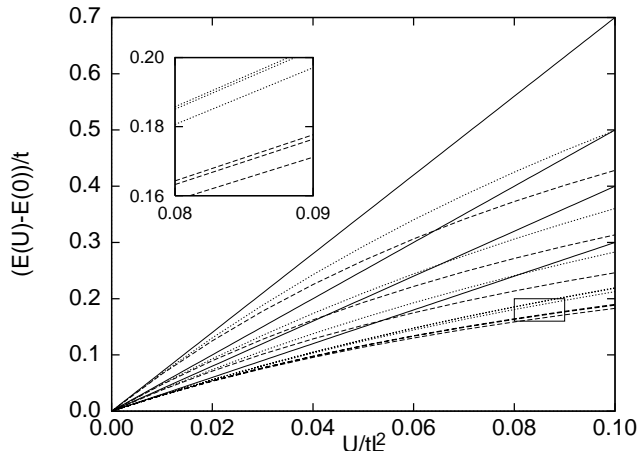}}
\caption{\label{fig:Perturbation}
The evolution of the 16-fold degenerated ground state multiplet
as a function of $U/L^2$. The exact numerical results for $L=5$
(dotted lines) and  $L=6$ (dashed lines) are contrasted with the
first order perturbation calculation (full straight lines). 
The insert is a magnification of that portion of the plot marked by
a rectangle showing how the exact calculation leads to a
splitting into three sub-multiplets with degeneracies 1, 2, and 4 of
the 7-fold degenerated perturbation theory ground state. The exact
ground state is nondegenerate. See also
Table~\protect\ref{tab:LowStates}.
}
\end{figure}

In Fig.~\ref{fig:Perturbation} we compare the exact numerical
calculation with Eq.~(\ref{eq:PertErg}) for $L = 5$ and 6. Note
especially how in the exact calculation the degeneracy of the
ground state energy is lifted completely. This is also demonstrated in
Table~\ref{tab:LowStates}, where it can be seen that also for $L=3$
the ground state is nondegenerate. However, the ground state of $L=4$
remains 3-fold degenerated  even in the exact calculation. This
reflects the particular symmetry of the $4\!\times\!4$ lattice,
where the hyperplane reflection $r_h$ leads to the existence of 3-fold
degenerate representations as shown in Appendix~\ref{sec:appendixA}.

Using the quantum numbers $(k_x,k_y,b_x,b_y,c)$ of
Table~\ref{tab:Irrep356} in Appendix~\ref{sec:appendixB} to interpret
the representation label $R$ in Table~\ref{tab:LowStates} we note that
they are of the same form for $L=3,5,6$: 
(0,0,1,1,-1), 
(0,0,1,-1,$*$),
($\frac{2\pi}{L}$,$\frac{2\pi}{L}$,$*$,$*$,-1),
($\frac{4\pi}{L}$,0,$*$,1,$*$),
($\frac{2\pi}{L}$,$\frac{2\pi}{L}$,$*$,$*$,1), and
(0,0,1,1,1), where the first set corresponds to the unique ground
state. This result extends to the $L$=4 lattice when the actual
space group is replaced by the one generated by
$\{t_x,r_x,t_y,r_y,r_d\}$. One state in the $L$=4 triplet ground state
belongs to the (0,0,1,1,-1) representation and the other two states
to ($\frac{4\pi}{L}$,0,1,1,$*$). In all cases
the ground state is odd with respect to the diagonal reflection $r_d$,
which is understandable since that symmetry suppresses the
ability of having pairs along the diagonal. 
Finally we note that due to the aforementioned spin symmetry of the
Hubbard model, the ground state of the $S_z$=0 sector is in fact a
global ground state, and furthermore since it has $S\!=\!0$ the same
energy level does not exist in any other $S_z$ sector. Thus we can
conclude that the global ground state in the generic case 
($L\neq4$) is nondegenerate. 

\newlength{\nl}
\settowidth{\nl}{$\! -1$}
\begin{table}[htb]
\begin{tabular}{
||c|c@{$\:\:$(}r@{,}r@{)$\:$}c||c@{$\:\:$(}r@{,}r@{,}r@{)$\:$}c||
    c@{$\:\:$(}r@{,}r@{)$\:$}c||c@{$\:\:$(}r@{,}r@{,}r@{)$\:$}c||}
& 
\multicolumn{4}{c||}{$L = 3$} & 
\multicolumn{5}{c||}{$L = 4$} & 
\multicolumn{4}{c||}{$L = 5$} & 
\multicolumn{5}{c||}{$L = 6$} 
\\ \hline 
\# 
& $E/t$ & $R$ & $S$       & $l_R$
& $E/t$ & $R$ & $S$ & $J$ & $l_R$
& $E/t$ & $R$ & $S$       & $l_R$
& $E/t$ & $R$ & $S$ & $J$ & $l_R$
\\ \hline \hline
P &  -9.94000 & -& -& 7&  -11.94000 &  -& -& -& 7
  & -13.17607 & -& -& 7&  -13.94000 & -& -& -& 7\\ \hline
0 &  -9.94122 & 2& 0& 1&  -11.94268 &  9& 0& 6& 3
  & -13.18059 & 2& 0& 1&  -13.94676 & 4& 0& 16& 1\\
1 &  -9.94110 & 8& 1& 4&  -11.94232 & 16& 1& 6& 6
  & -13.18012 & 4& 1& 2&  -13.94617 & 9& 1& 16& 2\\
2 &  -9.94097 & 4& 1& 2&  -11.90425 & 17& 0& 6& 6
  & -13.18005 &10& 1& 4&  -13.94606 &25& 1& 16& 4\\
3 &  -9.92188 & 5& 0& 4&  -11.86560 &  0& 0& 6& 1
  & -13.16204 & 7& 0& 4&  -13.92870 &14& 0& 16& 4\\
4 &  -9.90229 & 7& 0& 4&  \multicolumn{5}{c||}{$-$}
  & -13.14302 & 9& 0& 4&  -13.91032 &24& 0& 16& 4\\
5 &  -9.86204 & 0& 0& 1&  \multicolumn{5}{c||}{$-$}
  & -13.10590 & 0& 0& 1&  -13.87494 & 0& 0& 16& 1\\
\end{tabular}
\caption{\label{tab:LowStates} For $U/tL^2 = 0.02$ the 7-fold
degenerated perturbation theory ground state $E^{(1)}_{L,\beta=3}$
given in row $P$ is compared with the exact energies $E$ of the 16
levels splitting off from the $U\!=\!0$ ground state $E^{(0)}_L$.
The degeneracies of the sub-multiplets are given by the dimension
$l_R$ of the corresponding irreducible representation $R$ of the space
group. Note that except for $L$=4 the exact ground state is
nondegenerate. Also listed are the quantum numbers $(R,S,J)$ of the
levels.}
\end{table}

\subsection{Remaining degeneracies}
\label{subsec:RemainDeg}

After taking all the known symmetries into account and after
projecting into the symmetry invariant subspaces, it turns out that
within each subspace some further degeneracies remain. The reason 
is that the low filling of the lattice allows for a kind of restricted
permutation symmetry for the particle momentum components resulting in
energy eigenstates which are simultaneously eigenstates of $\hat{T}$
and $\hat{U}$ and therefore independent of $U$ (though
their eigenvalues might depend on $U$). We denote such states
$\hat{T}/\hat{U}$ states or $|\psi^{\gamma}_{S}\rangle$, where $S$ is
the spin and $\gamma$ the eigenvalue of $\hat{U}$. With four electrons
$\gamma$ takes the values $\gamma=0,1,2$ corresponding to
superposition of states containing exactly $\gamma$ pairs, i.e.\
doubly occupied sites. A generic energy eigenstate is not a
$\hat{T}/\hat{U}$ state and in that case we write $\gamma = *$. In
Table~\ref{tab:DegRemain} we show the number of generic energy
eigenstates and $\hat{T}/\hat{U}$ states found numerically. The
Hubbard model is thus partly integrable. The spectrum of each symmetry
invariant subspace is a mixture of an integrable component (the
$\hat{T}/\hat{U}$ states) and a non-integrable component (the generic
states), and in the analysis in Sec.~\ref{sec:SpectralStat} of the
spectral statistics we throw away the integrable component and analyze
only the generic non-integrable component.

The $\hat{T}/\hat{U}$ states are formed by specific superpositions
of eigenstates of $\hat{T}$ by permuting the eight momentum components
$k_n^{\zeta}$, $\zeta = x,y$ and $n = 0, 1, 2,$ and 3 such that the 
sum of cosines in Eq.~(\ref{eq:ZeroEigen}) remains unchanged.
In the following we mention some large classes of such states. The 
reader is referred to Appendix~\ref{sec:appendixD} for details.

\newlength{\LenMedium}
\setlength{\LenMedium}{0.70\textwidth}

\begin{table}[htb] 
\begin{center}
\begin{minipage}[t]{\LenMedium}
\begin{tabular}{|c|c||r@{/}l|r@{/}l|r@{/}l|r@{/}l|}
$\gamma$ & $S$ & 
\multicolumn{2}{c|}{$L=3$} & 
\multicolumn{2}{c|}{$L=4$} & 
\multicolumn{2}{c|}{$L=5$} & 
\multicolumn{2}{c|}{$L=6$} \\ \hline \hline
  & 0  & 540 & (0) & 4169 &   (0) & 31977 &    (0)& 115896 &     (0) \\
$*$&1  & 621 & (0) & 4472 &   (0) & 41662 &    (0)& 137199 &     (0) \\
  & 2  &   0 & (0) &    0 &   (0) &     0 &    (0)&      0 &     (0) \\
\hline \hline
  & 0  &   0 & (0) & 1176 &(1143) &   523 &  (316)&  23555 & (23427) \\
0 & 1  &   9 & (0) & 2548 &(2519) &  3188 & (2823)&  60306 & (60160) \\
  & 2  & 126 &(80) & 1820 &(1727) & 12650 &(12362)&  58905 & (58723) \\
\hline \hline
  & 0  &   0 & (0) &   94 &  (14) &     0 &    (0)&    408 &   (247) \\
1 & 1  &   0 & (0) &  120 &  (24) &     0 &    (0)&    630 &   (420) \\
  & 2  &   0 & (0) &    0 &   (0) &     0 &    (0)&      0 &     (0) \\
\hline \hline
  & 0  &   0 & (0) &    1 &   (0) &     0 &    (0)&      1 &     (0) \\
2 & 1  &   0 & (0) &    0 &   (0) &     0 &    (0)&      0 &     (0) \\
  & 2  &   0 & (0) &    0 &   (0) &     0 &    (0)&      0 &     (0) \\
\hline \hline
 & \% & 
\multicolumn{2}{c|}{89.6\%} &
\multicolumn{2}{c|}{60.0\%} &
\multicolumn{2}{c|}{81.8\%} &
\multicolumn{2}{c|}{63.8\%} \\
\end{tabular}
\end{minipage}
\end{center}

\vspace*{1mm}
\caption{\label{tab:DegRemain}
For  $L=3,4,5$, and 6 are shown the number of energy eigenstates found
numerically in each of the four main groups indexed by $\gamma$. The first
group, denoted $\gamma$=$*$, contains the generic energy eigenstates
that are not eigenstates of $\hat{U}$. 
The three other groups, denoted $\gamma = 0,1,2$ respectively, contain
the nongeneric $\hat{T}/\hat{U}$ states that are simultaneous
eigenstates of $\hat{U}$ (with eigenvalue $\gamma$) and $\hat{T}$. 
In parenthesis are given the number of states that  
remain degenerated in the symmetry invariant subspaces after taking
the space, spin, and pseudospin symmetries into account. Note how
the $\gamma$=$*$ states exhibit no further degeneracy. In the last row
denoted ``\%'' are given the percentages of $\gamma$=$*$ states out of
the total number of states.
}
\end{table}

First we note that for even $L$ proportionally many more
states remain degenerate as compared to $L$ odd. This difference is
due to the momentum component $k^{\zeta}_n = \pi$, which only exists
for $L$ even. Because $\cos(\pi\!-\!k) + \cos(k) = 0$ independent of
$k$ such terms drop out when $\hat{T}$ is applied to a state
containing this combination and a certain degree of freedom is left to
form the $\hat{T}/\hat{U}$ state superpositions. Starting with
two-pair states $\gamma=2$ we find exactly one energy eigenstate  
$|\psi_{S=0}^{\gamma=2} \rangle$. It depends on the momentum $\pi$,
and hence it exits only for even $L$. Similarly, for one-pair states
only even $L$ leads to $\hat{T}/\hat{U}$ states. A number
$(\stackrel{L^2}{_2})$ of states $|\psi_{S=1}^{\gamma=1} \rangle$ can
easily be constructed. More care must be taken upon forming
the corresponding $S$=0 states. However, we have succeeded in
constructing analytically the number of states $|\psi_{S=0}^{\gamma=1}
\rangle$ required by Table~\ref{tab:DegRemain}. Finally, for
zero-pair states the existence of the momentum $\pi$ is not required
to form the $\hat{T}/\hat{U}$ states, but it certainly helps. Thus
both even and odd values of $L$ lead to nongeneric energy
eigenstates, but relatively more such states are found for even $L$.
It is easy to see that all states with the maximal spin $S=2$ are
$\hat{T}/\hat{U}$ states $|\psi_{S=2}^{\gamma=0}\rangle$. This is a
trivial consequence of choosing four different sites (or momenta) and
forming the superposition given in Eq.~(\ref{eq:PS2}). There are 
$(\stackrel{L^2}{_4})$ such states. The construction of
states $|\psi_{S=S'}^{\gamma=0}\rangle$ with $S'$=1 or 0 is more
cumbersome, so in Appendix~\ref{sec:appendixD} we have only
given two examples of classes of such states.

The main result of this section is that the degeneracies that remain
after space, spin and pseudospin symmetry have been taking into
account, is related to a restricted permutation symmetry of the
momentum components. We have not found the associated projection
operators, but numerically and partly analytically we have established
the fact that the degenerate states are simultaneously eigenstates
of $\hat{T}$ and $\hat{U}$ with energies $E(U) = E(0) + \gamma U$. By
discarding these states we end up with nondegenerate
$\hat{U}$-dependent states. It is the spectral statistics of these
states we analyze in the following section, and this analysis confirm
our claim that all symmetries indeed have been taken into account.

\section{Spectral statistics of the Hubbard model}
\label{sec:SpectralStat}

Having sorted the spectrum according to all symmetries including the
restricted permutation symmetry of the momentum components the RMT
analysis can be performed. As discussed in Sec.~\ref{sec:RMT} the
first step is unfolding of the spectrum. There we mentioned how the
unfolding procedure is not uniquely determined, so we turn to that
problem first.

\subsection{Optimization of the unfolding procedure}
\label{subsec:OptimUnf}

\begin{figure}[h]
\epsfxsize=\textwidth \centerline{\epsfbox{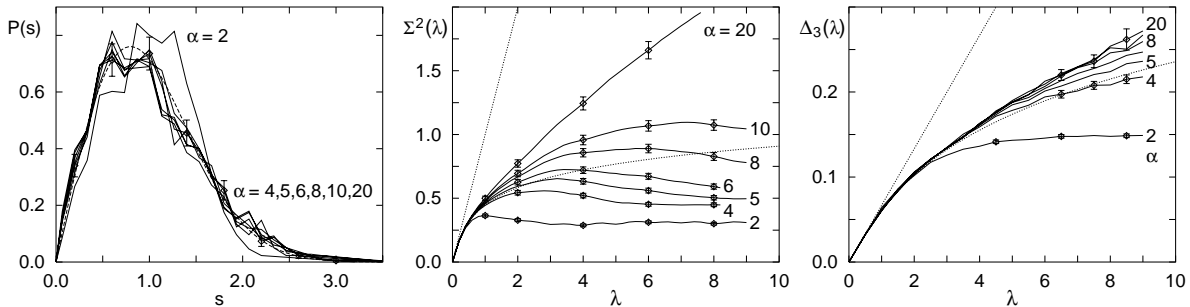}}
\caption{\label{fig:BestUnf}
The spectral statistics $P(s)$, $\Delta_3(\lambda)$, and
$\Sigma^2(\lambda)$ for the invariant subspace $(R,S)$ = $(13,1)$ of
the $5\!\times\!5$ lattice as a function of the free parameter
$\alpha$ of the unfolding procedure using Gaussian broadening with
variable width. In all three panels the values for $\alpha$ are 2, 4,
5, 6, 8, 10 and 20. In the first panel 6 of the 7 curves fluctuates
around  $P^{GOE}(s)$ hardly visible as a smooth dotted line. In the
two other panels the Poisson case are given as a dotted straight line
while the GOE case is given by a dotted smooth curve. For all data
sets a few representative error bars are shown.
}
\end{figure}

We unfold the spectra by using the method of Gaussian broadening with
variable width discussed in Sec.~\ref{subsec:unfolding}. The free
parameter $\alpha$ corresponding to how many energy levels each
Gaussian essentially spreads out over to each side, and the question
now arises how to choose it. The problem we face is illustrated in
Fig.~\ref{fig:BestUnf} were it is 
seen that although the level spacing distribution $P(s)$ is
essentially independent of the choice of $\alpha$, both the number
variance $\Sigma^2(\lambda)$ and the spectral rigidity
$\Delta_3(\lambda)$ varies with $\alpha$. It is seen that
$\Delta_3(\lambda)$ is less sensitive to changes in $\alpha$ than
$\Sigma^2(\lambda)$. The former seems to saturate for large values of
$\alpha$, while the latter continues to grow rapidly as $\alpha$
enhances. We have chosen that value $\alpha_0$ of $\alpha$ which makes
$\Sigma^2(\lambda)$ fit the corresponding  GOE curve as well as it
can. In general that leads to $\alpha_0 \approx 4$. We note that for
this choice of $\alpha$, $\Delta_3(\lambda)$ is  
almost saturated, while $P(s)$ remains unchanged. Thus, two of the
three statistics are essentially independent of variations of $\alpha$
around $\alpha_0$, while the third is as close to the GOE behavior as
it can be. 

\begin{figure}[h]
\epsfysize=60mm \centerline{\epsfbox{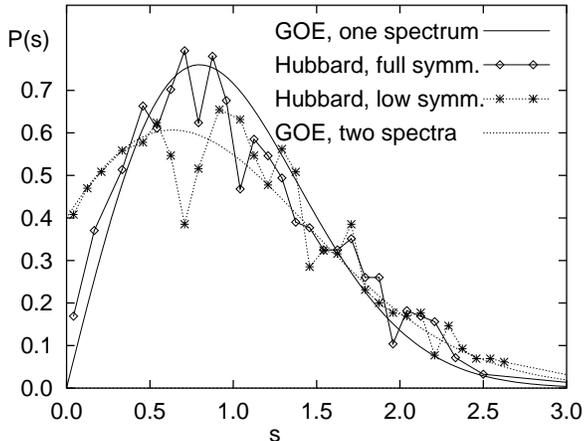}}
\caption{\label{fig:WrongSym}
$P(s)$ for $L=4$ with $U/t=10$ is calculated after sorting the spectra
using either the full symmetry group of the Hamiltonian ($\diamond$)
or an symmetry group artificially lowered ($*$) as described in the text.
The full symmetry case compares well with the Wigner surmise (smooth
full curve) whereas the lower symmetry case compares well with the
distribution of two GOE-spectra mixed with the relative weights 0.72
and 0.28 (dotted smooth curve).}
\end{figure}

To illustrate the importance of sorting the spectrum by group theory
we show in Fig.~\ref{fig:WrongSym} the level spacing distribution for
$L\!=\!4$ with $U/t=10$ for the case where all symmetries are taken
into account (full symm.) and for the case with lower symmetry
(low symm.) where the spin and pseudospin symmetries are being
kept intact but where  the space group has been reduced by replacing
among  the generators the special hyperplane 
reflection $r_h$ with the ordinary diagonal reflection $r_d$. For the
full symmetry case a distribution rather close to the Wigner surmise
is found, whereas the level repulsion is partially lost for the low
symmetry case, and the data fits reasonably well the distribution found by
mixing two GOE-spectra \cite{Pandey} with relative weights 0.72 and
0.28. This makes sense since by lowering the symmetry group 
artificially, spectra from the independent true symmetry invariant
subspaces are being mixed. For the more severe symmetry reduction
where the pseudospin is altogether neglected and the space group is
generated only by $\{t_x,t_y\}$, we find the level spacing
distribution to be the Poisson (exponential) distribution (not shown
in the figure). 

\subsection{The statistics $P(S)$, $\Sigma^2(\lambda)$, and
$\Delta_3(\lambda)$ of the Hubbard model}
\label{subsec:PofS}

In this subsection we present the results of the spectral statistical
analysis of the Hubbard model at low filling. We present only results
for $L$=4, 5, and 6 since $L$=3 yields too poor statistics due to its
small invariant subspaces. Besides letting $L$ vary we are also
varying the coupling strength $U/t$ and present results for weak,
intermediate and strong coupling regimes for $L=5$. To improve on the
statistics we have averaged over the largest invariant subspaces.
For $P(s)$ the size $\delta P$ of the error bars shown in the figures
are estimated by $\delta P_i = C \sqrt{n_i}/h_i$, where $n_i$ is the
number of points in bin $i$ of the associated histogram, $h_i$ is the
width of the bin, while $C$ is the normalization factor rendering a
total probability of 1. For $\Sigma^2(\lambda)$ and
$\Delta_3(\lambda)$ the error bars are estimated by ordinary variances 
obtained from the values calculated in the many contiguous intervals
of length $\lambda$ throughout the unfolded spectrum.

\begin{figure}[h]
\epsfysize=60mm \centerline{\epsfbox{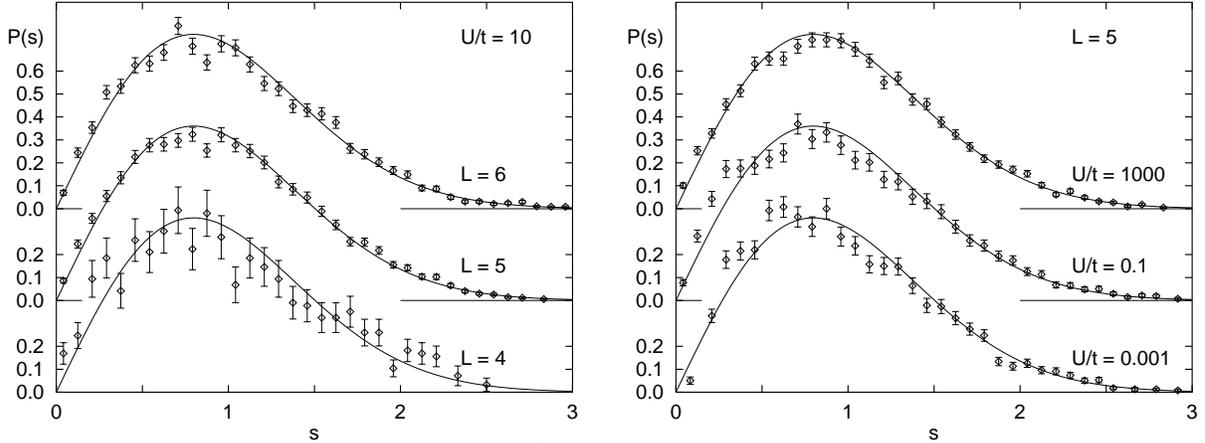}}
\caption{\label{fig:PdS}
The probability distribution $P(s)$ of the level spacings $s$ averaged
over the largest symmetry invariant subspaces for the Hubbard model
with four electrons on square $L\!\times\!L$ lattices. To the left is
shown $P(s)$ as a function of lattice size $L$ at medium coupling
strength. The data represents $L$=4, 5, and 6 for $U/t=10.0$. To the
right is shown $P(s)$ as a function of coupling strength at fixed low
filling. The data represents $U/t$=0.1, 10, and 1000 for $L$=5. The
full line is the Wigner distribution found for GOE random 
matrices.}  
\end{figure}

The results for $P(s)$ are shown in Fig.~\ref{fig:PdS}. It is seen
that for all lattice sizes and for any value of the coupling strength
the level spacing distribution is fairly close to the Wigner
distribution of GOE; it possesses a pronounced linear level repulsion
for small $s$, a peak near $s=0.8$ signaling spectral rigidity, and a
rapid fall off for $s>2$.

\begin{figure}[h]
\epsfysize=60mm \centerline{\epsfbox{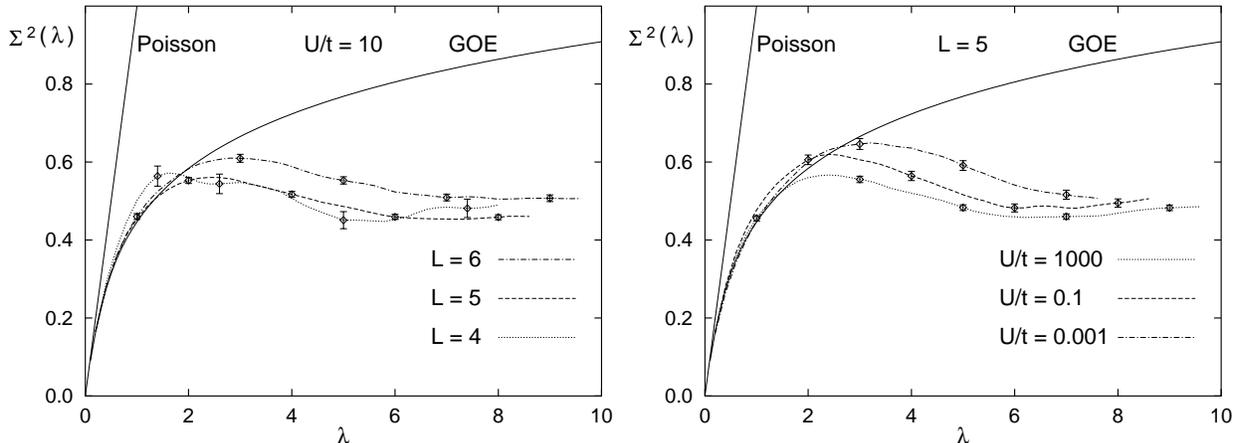}}
\caption{\label{fig:Sigma2}
The number variance $\Sigma^2(\lambda)$ calculated for the same
parameters as in Fig.~\protect\ref{fig:PdS}. The data are compared to
the results of the random diagonal matrix ensemble (Poisson) and the
random full matrix ensemble (GOE), shown as the straight full line and
the curved full line, respectively.
}
\end{figure}

In Fig.~\ref{fig:Sigma2} is shown $\Sigma^2(\lambda)$ of the Hubbard
model for the same parameters as for $P(s)$ just mentioned. When
$\lambda$ is small the rigidity of the Hubbard spectrum is very close
to that of the GOE random matrices, while for larger $\lambda$ a
saturation sets in. For all values of $U/t$ we find the critical
value $\lambda^*$ where the departure from GOE sets in to be roughly
2. The precise origin of $\lambda^*$ remains unclear.

\begin{figure}[h]
\epsfysize=60mm \centerline{\epsfbox{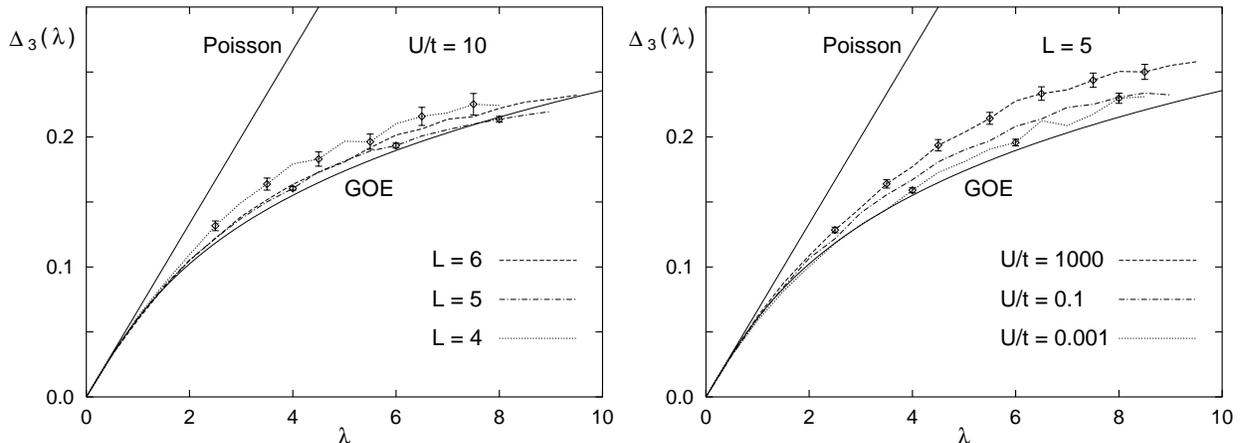}}
\caption{\label{fig:Delta3}
The spectral rigidity $\Delta_3(\lambda)$ calculated for the same
parameters as in Fig.~\protect\ref{fig:PdS}. The data are compared to
the results of the random diagonal matrix ensemble (Poisson) and the
random full matrix ensemble (GOE), shown as the straight full line and
the curved full line, respectively.
}
\end{figure}

Finally, in Fig.~\ref{fig:Delta3} are shown the results for the spectral
rigidity $\Delta_3(\lambda)$. As $\Sigma^2(\lambda)$ also
$\Delta_3(\lambda)$ displays an excellent agreement with GOE for 
$\lambda < \lambda^* \approx 2$, for all fillings and for all values of
$U/t$. The deviations from GOE beyond $\lambda^*$ are not so marked.
The curves lie between the Poisson line and the GOE curve, but rather
close to the latter.

It is remarkable how the results for the three statistics studied are
fairly independent of the size of the lattice (equivalent to the
filling) and of the coupling strength. We find GOE-like behavior
not only for all finite values of $U/t$ including those close to the 
integrable $U=0$ limit, but also for filling factors as low as 0.06
close to the integrable single-particle limit.  However, as is evident
from the behavior of especially $\Sigma^2(\lambda)$ at large energy
scales, the Hubbard model cannot be modeled exactly by a simple GOE
random matrix model.

\section{Conclusions and discussion}
\label{sec:}

In this paper the energy level statistics of the Hubbard model for
$L\!\times\!L$ square lattices ($L=3,4,5,6$) at low filling (four
electrons) has been studied numerically for a wide range of the
coupling strength. 
With great care all known symmetries of the model (space,
spin and pseudospin symmetry) have been taken into account explicitly
from the beginning of the calculation by projecting into symmetry
invariant subspaces. The details of this group theoretical treatment
was presented with special attention to the nongeneric case of $L=4$,
where a particular complicated space group appears. The resulting
reduction of the numerical diagonalization is significant, and the
method presented can in a straightforward manner be extended to larger
 lattices and higher fillings and thus form the basis of improved
numerical studies of the Hubbard model and related models without
disorder. In particular, this work can be used as a starting point for
calculating various spectral functions, for which explicit forms of
the eigenstates are required. This will be dealt with in forthcoming
work. 

For all the lattices studied a significant amount of levels within
each symmetry invariant subspace remain degenerated, but except for
$L=4$ the ground state is nondegenerate. We explained the degenerate
states as a consequence of a restricted permutation symmetry of the
momentum components. These states, all independent of $U$, form an
integrable part of the spectrum, and after discarding them we end up
with nondegenerate spectra on which the level statistical 
analysis could be performed. 

The intricate structure of the Hubbard spectra necessitated the
development of a careful unfolding procedure as a preparatory step
before the level statistical analysis. The procedure we arrived
at tested favorably in many cases of pathological spectra, and it
seems to be very robust and applicable in general cases were no other 
natural unfolding procedure exist.

Finally, we have performed a level statistical analysis of the Hubbard
spectra, and we presented  results for the level spacing distribution
$P(s)$, the number variance $\Sigma^2(\lambda)$, and the spectral 
rigidity $\Delta_3(\lambda)$. The statistics for the different lattice
sizes and for a wide range of coupling strengths are essentially the
same: $P(s)$ shows a good agreement with GOE.
$\Sigma^2(\lambda)$  agrees only with GOE up to the $U/t$-independent
medium sized energy scale $\lambda^* \approx 2$ beyond which a
saturation sets in. $\Delta_3(\lambda)$ also agrees with GOE for
$\lambda < \lambda^*$. The deviation from GOE beyond $\lambda^*$ is
not so marked as that for $\Sigma^2(\lambda)$. The curve falls between
that of GOE and Poisson but rather close to the former.  
We stress that these results were also obtained for very
small coupling strengths approaching the integrable zero coupling
limit. This emphasizes the nonperturbative nature of the model
revealed by our analysis: even the smallest deviation from the
integrable limits leads to spectral statistics usually associated with
non-integrability and quantum chaos, and in this sense the model seems
always to be in the strong coupling limit.

Largely, our results show GOE-behavior of the spectral
statistics of the typical high lying excitations of the Hubbard model
at low filling. This indicates that at least the incoherent part of the
electronic spectral functions (related to the coherent part describing
the low lying electronic excitations through sum rules) are out of
reach by standard methods. On the other hand, it should be possible to
model this part by a random matrix ansatz. This is in agreement with
previous results of the $tJ$ model near half filling \cite{Montambaux}.
However, the cause of the deviations from GOE we found in
$\Sigma^2(\lambda)$ and $\Delta_3(\lambda)$ beyond $\lambda^*$ remains
an open question. The similar question has been answered in general
for single-particle systems with mean-level spacing $\Delta$: in
disordered (metallic) systems $\lambda^* \Delta \sim \hbar/\tau_D$,
where $\tau_D$ is the time it takes a particle to diffuse through the
system \cite{Altshuler}, whereas for pure (ballistic) systems
$\lambda^* \Delta \sim \hbar/\tau_0$, where $\tau_0$ is the period of
the shortest periodic orbit \cite{Berry2}. Results are also beginning
to emerge for disordered interacting lattice systems, where
$\lambda^*$ is related to the ratio $U/W$ between some interaction
strength $U$ and the disorder induced single-particle band width $W$
\cite{Berkovits,Weinmann}. For these systems the deviations from GOE 
are due to the preferential basis supplied by the given disorder
potential. For example the system consisting of two interacting
particles in a disorder potential \cite{Weinmann} could be studied
analytically by adding a random diagonal matrix modeling the disorder
single-particle states to a random GOE matrix modeling the
interactions between these states, and it was found that
$\Sigma^2(\lambda)$ increased as a 
power law for $\lambda > \lambda^*$. This behavior is in contrast to
the saturation we found (see Fig.~\ref{fig:Sigma2}), which looks more
like the result of the ballistic single-particle case \cite{Berry2}.
It is perhaps not surprising that such a similarity exists between the 
disorder-free chaotic single-particle case and the disorder-free Hubbard model
rather than between two strongly correlated systems one with and the
other without disorder. However, exactly what physical mechanism
produces a preferential basis for the Hubbard model at low filling is
not known, and neither is it known why a similar mechanism is
suppressed for the $tJ$ model near half filling, where much less
pronounced deviations from GOE are found \cite{Montambaux}.
These questions are topics for future work.

\section{Acknowledgements}
It is a pleasure to thank Benoit Dou\c{c}ot, Jean-Louis Pichard,
Cl\'ement Sire and Dietmar Weinmann for stimulating discussions and
Elliot H.~Lieb for valuable comments. We thank the Centre Grenoblois
de Calcul Vectoriel for its kind help with the part of our numerical
work that was carried out at its CRAY-94 computer.  H.B.\ was supported
by the European Commission under grant no.\ ERBFMBICT 950414.

\appendix

\section{The irreducible representations of $G_4$}
\label{sec:appendixA}
In Sec.~\ref{subsec:SpaceSymm} we showed how the group $G_4$ of
neighbor-conserving transformation of the $4\!\times\!4$ lattice is
isomorphic with the point group of the four-dimensional hypercube. In
this appendix we determine the structure of this group and we sketch
how the irreducible representations are found analytically. 
The fundamental simplification is the observation that $G_4$ has the
structure of a semi-direct product involving an invariant Abelian
subgroup. The theorem of induced representations\cite{Cornwell} can
then be used to find all irreducible representation. The theorem is
stated below. The coordinate set of a corner in the unit hypercube in
four dimensions is given as a four bit binary number. Any
transformation of the hypercube can thus be written as 
a permutation of the four bits followed by bit-inversion ($0
\leftrightarrow 1$) of all, some or none of the bits. 

In what follows any quadruple $(x_1,x_2,x_3,x_4)$ is written
in short hand notation as $(x_i)$. The group of bit permutations is
of course the permutation group $S_4$ denoted $\cal B$ in the
following to be consistent with Ref.~\onlinecite{Cornwell}, on which
the group theoretical work in this appendix is based. Any
element $b\!\in\!{\cal B}$ is written as $b = (b_i) = (b_1,b_2,b_3,b_4)$,
listing the permutation of the numbers 1, 2, 3, and 4. The group of
bit inversions is denoted $\cal A$. It is easily seen that ${\cal A} = 
{\cal Z}_2 \otimes {\cal Z}_2 \otimes {\cal Z}_2 \otimes {\cal Z}_2$,
since bit-inversions does or does not take place on each of the four
bit positions. Any element $a\!\in\!{\cal A}$ has the form $a = (a_i) =
(a_1,a_2,a_3,a_4)$, with $a_i = \pm 1$, where $+1$ means no
bit-inversion and $-1$ means bit-inversion. Any element of 
$g\!\in\!G_4$ can be written as $g = ab = (a_ib_i)$ and conversely all
products $ab\!\in\!G_4$. $\cal A$ contains 16 elements and
$\cal B$ contains 24 so that $G_4$ contains 384 elements. It is readily
verified that $\cal A$ is an Abelian subgroup of $G_4$. Furthermore,
$\cal A$ is an invariant subgroup since for $\forall a \!\in\!
{\cal A}, \forall b\!\in\!{\cal B}$: $bab^{-1} = b(a_i)b^{-1} =
(a_{b(i)})bb^{-1} = (a_{b(i)})\!\in\!{\cal A}$. Finally, the
only common element of $\cal A$ and $\cal B$ is the identity. We can
therefore conclude that $G_4$ is a semi-direct product of the invariant
Abelian subgroup $\cal A$ with $\cal B$:

\begin{equation}
G_4 = {\cal A} \semidir {\cal B} = ({\cal Z}_2 \!\otimes\! {\cal Z}_2
\!\otimes\! {\cal Z}_2 \!\otimes\! {\cal Z}_2) \semidir S_4.
\end{equation}

The first step in calculating the irreducible representations for
$G_4$ is to construct the character table $\chi^{\bf q}(a)$ of $\cal
A$. This is easily found as the product of the character table {\tiny
\begin{tabular}{|rr} \hline 1&1 \\ 1&-1 \end{tabular}} for ${\cal
Z}_2$ with itself four times. The 16 irreducible representations are
identified by the index ${\bf q} = 
(q_1,q_2,q_3,q_4)$, with $q_i = 0 \; \mbox{or} \; 1$, 
and the $\bf q$-th character
$\chi^{\bf q}(a)$ is given by the  product of $a_i$ to the power
$q_i$:

\begin{equation}
\label{eq:chia}
\chi^{\bf q}(a) = \prod_{i=1}^4 a_i^{q_i}.
\end{equation}
Next step is to pick any $\bf q'$ and construct the associated
little group ${\cal B}({\bf q}') \subseteq {\cal B}$ defined as

\begin{equation}
\label{eq:Bq}
{\cal B}({\bf q}') \equiv \left\{ 
b\!\in\!{\cal B} \mid \chi^{\bf q'}(bab^{-1}) = \chi^{\bf q'}(a),
\; \mbox{for} \; \forall a\!\in\!{\cal A}
\right\}.
\end{equation}
The group $\cal B$ is then written in a coset
decomposition after ${\cal B}({\bf q}')$:

\begin{equation}
{\cal B} = {\cal B}({\bf q}')b_1 \oplus {\cal B}({\bf q}')b_2 \oplus
\ldots \oplus {\cal B}({\bf q}')b_{M'}.
\end{equation}
For each of the $M'$  coset representatives $b_j$ an index ${\bf q'}_j$
is determined such that

\begin{equation}
\chi^{{\bf q'}_j} = \chi^{\bf q'}(b_jab_j^{-1}), \; \mbox{for} \;
\forall a\!\in\!{\cal A}.
\end{equation}
Note that $b_1$ is the identity and that ${\bf q'}_1$ hence equals
$\bf q'$. The set $\rm{orb}({\bf q'}) = \{ 
{\bf q'}_1, {\bf q'}_2, \ldots ,{\bf q'}_{M'}\}$ is called the orbit
of $\bf q`$. Now pick a $\bf q''$ outside orb($\bf q'$) and repeat
the procedure. This is continued until each of the 16 {\bf q}'s are
associated with an orbit. The last thing to do before constructing the
irreducible representations of $G_4$ is to find the irreducible
representations $\Delta^{{\bf q'}p}$ of the little group ${\cal B}({\bf
q'})$. This is usually a simple step due to the small size of ${\cal
B}({\bf q'})$. 

The theorem of induced representations\cite{Cornwell} states that
all irreducible representations $\Gamma^{{\bf q'}p}$ of the
semi-direct product ${\cal A} \semidir {\cal B}$, $\cal A$ being an
invariant Abelian subgroup, are found as follows. 
$(i)$ Pick one $\bf q'$ from each orbit. 
$(ii)$ Construct of the little group ${\cal B}({\bf q'})$ and its 
$n'_p$ irreducible representations $\Delta^{{\bf q'}p},
p=1,\ldots,n'_p$. 
$(iii)$ Find the $M'$ coset representatives $b_j\!\in\!{\cal
B},j=1,\ldots,M'$ of $\cal B$ with respect to ${\cal B}({\bf q'})$.
$(iv)$ Then the matrix elements of $\Gamma^{{\bf q'}p}$
for element $ab$ is given by:

\begin{equation}
\label{eq:G4irrep}
\Gamma^{{\bf q'}p}(ab)_{kt,jr} = \left\{
\begin{array}{ll}
\chi^{\bf q'}(a)
\left[ \Delta^{{\bf q'}p}(b_kbb_j^{-1})\right]_{tr} & \; \mbox{if} \;
b_kbb_j^{-1}\!\in\!{\cal B}({\bf q'}) \\
0   & \; \mbox{if} \; b_kbb_j^{-1} \not\in {\cal B}({\bf q'}). 
\end{array} \right.
\end{equation}

Here we will not give the explicit expressions of the irreducible
representations of $G_4$, but rather just briefly sketch the
construction of them and calculate how many there are and what
is the dimension of each of them.

First we chose ${\bf q'} = (0000)$. From Eq.~(\ref{eq:chia}) it is
easily seen that $\chi^{(0000)}(a) = 1$ for $\forall a\!\in\!{\cal A}$.
Hence $\chi^{(0000)}(bab^{-1}) = \chi^{(0000)}(a)$ for $\forall a \in
{\cal A}, \forall b\!\in\!{\cal B}$, and according to Eq.~(\ref{eq:Bq})
we find ${\cal B}(0000) = {\cal
B}$. The coset representation of $\cal B$ consists of only one term
and the single coset representative $b_1$ is the identity. As a
consequence we have $\mbox{orb}(0000) = \{(0000)\}$. Finally,
the irreducible representations $\Delta^{0p}$ of ${\cal B}(0000)$ are
simply those of ${\cal B} = S_4$ (or $T_d$ as the group also is called
\cite{Cornwell}), i.e.\ there are 5 different $\Delta^{0p}$-matrices
with dimensions 1, 1, 2, 3, and 3 respectively. From
Eq.~(\ref{eq:G4irrep}) we find that $j,k = 1$ and thus we have found 5
irreducible representations of $G_4$ with dimensions 1, 1, 2, 3, and 3
of the form $\Gamma^{(0000)p}(ab) = \chi^{(0000)}(a) \Delta^{0p}(b)$.

Next we choose ${\bf q'} = (1111)$. We find that for $\forall a \in
{\cal A}, \forall b\!\in\!{\cal B}$: $\chi^{(1111)}(bab^{-1}) = \prod_i
a^1_{b(i)} = \prod_j a_j = \chi^{(1111)}(a)$. So in analogy with ${\bf
q'} = (0000)$ we have ${\cal B}(1111) = {\cal B}$, and like before we
find 5 irreducible representations of $G_4$ with dimensions 1, 1, 2,
3, and 3 of the form $\Gamma^{(1111)p}(ab) = \chi^{(1111)}(a)
\Delta^{0p}(b)$, with the same $\Delta^{0p}$-matrices but different
$\chi$-prefactors as for ${\bf q'} = (0000)$.

We go on with ${\bf q'} = (1000)$. This yields
$\chi^{(1000)}(bab^{-1}) = a_{b(1)}$ which equals $\chi^{(1000)}(a)$
for $\forall a\!\in\!{\cal A}$ if and only if $b(1) = 1$. Thus ${\cal
B}(1000)$ is the six-element subgroup of $\cal B$ which leaves the
first axis invariant. The coset decomposition of $\cal B$ contains
four terms with coset representatives that each leaves one of the four
axis invariant. Not surprisingly we find $\mbox{orb}(1000) = \{
(1000), (0100), (0010), (0001) \}$. The little group ${\cal B}(1000)$
is isomorphic with $C_{3v}$ and has thus 3 irreducible representations
$\Delta^{1p}$ with dimensions 1, 1, and 2 respectively. This combined
with the fact that index $j,k$ in Eq.~(\ref{eq:G4irrep}) runs over the
four coset representatives means that we have found 3 more irreducible
representations $\Gamma^{(1000)p}$ of $G_4$ with dimensions 4, 4, and
8, and with entries of the form 0,
$\chi^{(1000)}\Delta^{1p}$, $\chi^{(0100)}\Delta^{1p}$, 
$\chi^{(0010)}\Delta^{1p}$, or $\chi^{(0001)}\Delta^{1p}$.

\begin{table}[t]
\begin{center}
{\small
\begin{tabular}{|c||rrrrrrrrrrrrrrrrrrrr|} 
R\verb+\+C & 0 & 1 & 2 & 3 & 4 & 5 & 6 & 7 & 8 & 9 & 10 & 11 & 12 & 13 & 14 
& 15 & 16 & 17 & 18 & 19 \\ \hline \hline 
0 & 1 & 1 & 1 & 1 & 1 & 1 & 1 & 1 & 1 & 1 & 1 & 1 & 1 & 1 & 1 & 1 & 1
& 1 & 1 & 1 \\ 
1 & 1 & 1 & 1 & 1 & 1 & 1 & 1 & 1 & 1 & 1 & 1 & -1 & -1 & -1 & -1 & -1
& -1 & -1 & -1 & -1 \\ 
2 & 1 & 1 & 1 & 1 & 1 & 1 & 1 & -1 & -1 & -1 & -1 & 1 & 1 & 1 & 1 & -1
& -1 & -1 & -1 & -1 \\ 
3 & 1 & 1 & 1 & 1 & 1 & 1 & 1 & -1 & -1 & -1 & -1 & -1 & -1 & -1 & -1
& 1 & 1 & 1 & 1 & 1 \\ \hline
4 & 2 & 2 & 2 & 2 & 2 & -1 & -1 & 0 & 0 & 0 & 0 & 0 & 0 & 0 & 0 & 1 &
1 & -2 & -2 & -2 \\ 
5 & 2 & 2 & 2 & 2 & 2 & -1 & -1 & 0 & 0 & 0 & 0 & 0 & 0 & 0 & 0 & -1 &
-1 & 2 & 2 & 2 \\ \hline 
6 & 3 & 3 & -1 & 3 & -1 & 0 & 0 & 1 & 1 & -1 & 1 & -1 & -1 & 1 & -1 &
0 & 0 & 1 & -3 & -3 \\ 
7 & 3 & 3 & -1 & 3 & -1 & 0 & 0 & -1 & -1 & 1 & -1 & 1 & 1 & -1 & 1 &
0 & 0 & 1 & -3 & -3 \\ 
8 & 3 & 3 & -1 & 3 & -1 & 0 & 0 & -1 & -1 & 1 & -1 & -1 & -1 & 1 & -1
& 0 & 0 & -1 & 3 & 3 \\  
9 & 3 & 3 & -1 & 3 & -1 & 0 & 0 & 1 & 1 & -1 & 1 & 1 & 1 & -1 & 1 & 0
& 0 & -1 & 3 & 3 \\ \hline 
10 & 4 & -4 & 0 & 0 & 0 & 1 & -1 & -2 & 2 & 0 & 0 & 2 & -2 & 0 & 0 &
-1 & 1 & 0 & -2 & 2 \\  
11 & 4 & -4 & 0 & 0 & 0 & 1 & -1 & 2 & -2 & 0 & 0 & -2 & 2 & 0 & 0 &
-1 & 1 & 0 & -2 & 2 \\ 
12 & 4 & -4 & 0 & 0 & 0 & 1 & -1 & 2 & -2 & 0 & 0 & 2 & -2 & 0 & 0 & 1
& -1 & 0 & 2 & -2 \\ 
13 & 4 & -4 & 0 & 0 & 0 & 1 & -1 & -2 & 2 & 0 & 0 & -2 & 2 & 0 & 0 & 1
& -1 & 0 & 2 & -2 \\ \hline
14 & 6 & 6 & 2 & -2 & -2 & 0 & 0 & 0 & 0 & 0 & 0 & -2 & -2 & 0 & 2 & 0
& 0 & 0 & 0 & 0 \\ 
15 & 6 & 6 & -2 & -2 & 2 & 0 & 0 & -2 & -2 & 0 & 2 & 0 & 0 & 0 & 0 & 0
& 0 & 0 & 0 & 0 \\ 
16 & 6 & 6 & -2 & -2 & 2 & 0 & 0 & 2 & 2 & 0 & -2 & 0 & 0 & 0 & 0 & 0
& 0 & 0 & 0 & 0 \\ 
17 & 6 & 6 & 2 & -2 & -2 & 0 & 0 & 0 & 0 & 0 & 0 & 2 & 2 & 0 & -2 & 0
& 0 & 0 & 0 & 0 \\ \hline
18 & 8 & -8 & 0 & 0 & 0 & -1 & 1 & 0 & 0 & 0 & 0 & 0 & 0 & 0 & 0 & 1 &
-1 & 0 & -4 & 4 \\  
19 & 8 & -8 & 0 & 0 & 0 & -1 & 1 & 0 & 0 & 0 & 0 & 0 & 0 & 0 & 0 & -1
& 1 & 0 & 4 & -4 \\ 
\end{tabular}
} 
\end{center}
\caption{\label{tab:Chi4}
The character table of $G_4$ defining the index $R$ for each of the 20
irreducible representations and the index $C$ for each of the 20
classes. Column $C=0$ contains the dimension $l_R$ ranging from 1 to 8
of the representations.}
\end{table}

Then we consider ${\bf q'} = (0111)$. The character $\chi^{\bf q'}$
now gives $\chi^{(0111)}(bab^{-1}) = a_{b(2)}a_{b(3)}a_{b(4)}$ which
as above equals $\chi^{(0111)}(a)$ for $\forall a\!\in\!{\cal A}$ if and
only if $b(1) = 1$. The rest of the analysis is similar as the
previous case: $\mbox{orb}(0111) = \{(0111), (1011), (1101), (1110)
\}$ and ${\cal B}(1000)$ is isomorphic with $C_{3v}$. We end up with 3
more irreducible representations $\Gamma^{(0111)p}$ of $G_4$ with
dimensions 4, 4, and 8, having entries of 0,
$\chi^{(0111)}\Delta^{1p}$, $\chi^{(1011)}\Delta^{1p}$,
$\chi^{(1101)}\Delta^{1p}$, or $\chi^{(1110)}\Delta^{1p}$. 

The last choice for $\bf q'$ turns out to be ${\bf q'} = (0011)$, and
we obtain $\chi^{(0011)}(bab^{-1}) = a_{b(3)}a_{b(4)}$, which equals
$\chi^{(0011)}(a)$ for $\forall a\!\in\!{\cal A}$ if and only if 
$b$ does not mix the pairs (1,2) with (3,4). It is easily seen that ${\cal
B}(0011)$ is a 4-element group isomorphic with ${\cal Z}_2 \otimes
{\cal Z}_2$, and thus has 4 1-dimensional irreducible representations
$\Delta^{2p}$. The coset representation of $\cal B$ consists of 6
terms in this case, and $\mbox{orb}(0011) = \{ (0011), (0101), (0110),
(1001), (1010), (1100) \}$. At this point we note that all 16 possible
values of $\bf q'$ now is a member of an orbit. The 6 coset
representatives combined with the 4 1-dimensional irreducible
representations of ${\cal B}(0011)$ yields through Eq.~(\ref{eq:G4irrep}) 4
new irreducible representations of $G_4$ each being 6-dimensional and
each having entries of the form 0,
$\chi^{(1100)}\Delta^{2p}$, $\chi^{(1010)}\Delta^{2p}$, 
$\chi^{(1001)}\Delta^{2p}$, $\chi^{(0110)}\Delta^{2p}$, 
$\chi^{(0101)}\Delta^{2p}$, or $\chi^{(0011)}\Delta^{2p}$.
In conclusion we see that $G_4$ has 20 irreducible representations
with the dimensions listed in the first column of the character table
shown in Table~\ref{tab:Chi4}. This result is to be contrasted with the
list of representation dimensions given in Ref.~\onlinecite{Fano},
where only translations and reflections are taken into account in the
analysis of the space group of the $4 \!\times\! 4$ lattice. Note
especially the 3- and 6-dimensional representations found here as
opposed to dimensions equal powers of 2 in Ref.~\onlinecite{Fano}.

\section{The irreducible representations of $G_3$, $G_5$, and $G_6$}
\label{sec:appendixB}

\vspace*{-5mm}

\begin{table}[h]

\newlength{\LenSmall}
\setlength{\LenSmall}{0.39\textwidth}
\newlength{\LenBig}
\setlength{\LenBig}{0.58\textwidth}

\begin{minipage}[t]{\LenSmall}
\begin{tabular}{||c|cccc|c|cccc||} 
\multicolumn{9}{||c}{The $3\!\times\!3$ lattice}&\\
\hline \hline
$R$&0&1&2&3&4&5&6&7&8\\
\hline
$k_x$&0&0&0&0&0&
$\frac{2\pi}{3}$&$\frac{2\pi}{3}$&$\frac{2\pi}{3}$&$\frac{2\pi}{3}$\\
$k_y$&0&0&0&0&0&0&0&$\frac{2\pi}{3}$&$\frac{2\pi}{3}$\\
$b_x$&1&--1&1&--1&1&$*$&$*$&$*$&$*$\\
$b_y$&1&--1&1&--1&--1&1&--1&$*$&$*$\\
$c$&1&1&--1&--1&$*$&$*$&$*$&1&--1\\ 
\hline
$l_R$&1&1&1&1&2&4&4&4&4\\
\end{tabular}
\end{minipage}
\hfill
\begin{minipage}[t]{\LenBig}
\begin{tabular}{||c|cccc|c|cccccccc|c||} 
\multicolumn{14}{||c}{The $5\!\times\!5$ lattice}&\\
\hline \hline
$R$&0&1&2&3&4&5&6&7&8&9&10&11&12&13\\
\hline
$k_x$&0&0&0&0&0&
$\frac{2\pi}{5}$&$\frac{2\pi}{5}$&$\frac{4\pi}{5}$&$\frac{4\pi}{5}$&
$\frac{2\pi}{5}$&$\frac{2\pi}{5}$&$\frac{4\pi}{5}$&$\frac{4\pi}{5}$&
$\frac{2\pi}{5}$\\
$k_y$&0&0&0&0&0&0&0&0&0&
$\frac{2\pi}{5}$&$\frac{2\pi}{5}$&$\frac{4\pi}{5}$&$\frac{4\pi}{5}$&
$\frac{4\pi}{5}$\\
$b_x$&1&--1&1&--1&1&$*$&$*$&$*$&$*$&$*$&$*$&$*$&$*$&$*$\\
$b_y$&1&--1&1&--1&--1&1&--1&1&--1&$*$&$*$&$*$&$*$&$*$\\
$c$&1&1&--1&--1&$*$&$*$&$*$&$*$&$*$&1&--1&1&--1&$*$\\ 
\hline
$l_R$&1&1&1&1&2&4&4&4&4&4&4&4&4&8\\
\end{tabular}
\end{minipage}

\vspace*{2mm}

\begin{tabular}{||c|cccccccc|cccccc|cccccccccccc|c||} 
\multicolumn{27}{||c}{The $6\!\times\!6$ lattice}&\\
\hline \hline
$R$&0&1&2&3&4&5&6&7&8&9&10&11&12&13&14&15&16&17&18&19&20&21&22&23&24&25&26\\
\hline
$k_x$&0&$\pi$&0&$\pi$&0&$\pi$&0&$\pi$&$\pi$&0&$\pi$&$\pi$&$\pi$&$\pi$&
$\frac{2\pi}{3}$&$\frac{2\pi}{3}$&$\frac{\pi}{3}$&$\frac{\pi}{3}$&
$\frac{\pi}{3}$&$\frac{\pi}{3}$&$\frac{2\pi}{3}$&$\frac{2\pi}{3}$&
$\frac{2\pi}{3}$&$\frac{2\pi}{3}$&$\frac{\pi}{3}$&$\frac{\pi}{3}$&
$\frac{\pi}{3}$\\
$k_y$&0&$\pi$&0&$\pi$&0&$\pi$&0&$\pi$&0&0&0&$\pi$&0&0&0&0&$\pi$&$\pi$&
0&0&$\pi$&$\pi$&$\frac{2\pi}{3}$&$\frac{2\pi}{3}$&
$\frac{\pi}{3}$&$\frac{\pi}{3}$&$\frac{2\pi}{3}$\\
$b_x$&1&--1&--1&1&1&--1&--1&1&--1&1&1&1&--1&1&
 $*$&$*$&$*$&$*$&$*$&$*$&$*$&$*$&$*$&$*$&$*$&$*$&$*$\\
$b_y$&1&--1&--1&1&1&--1&--1&1&1&--1&1&--1&--1&--1&1&--1&--1&1&1&--1&--1&1&
 $*$&$*$&$*$&$*$&$*$\\
$c$&1&1&1&1&--1&--1&--1&--1&
 $*$&$*$&$*$&$*$&$*$&$*$&$*$&$*$&$*$&$*$&$*$&$*$&$*$&$*$&1&--1&1&--1&$*$\\
\hline
$l_R$&1&1&1&1&1&1&1&1&2&2&2&2&2&2&4&4&4&4&4&4&4&4&4&4&4&4&8\\
\end{tabular}

\caption{\label{tab:Irrep356}
The irreducible representations $R$ of $G_L$ for $L$ = 3, 5, and 6,
their corresponding quantum numbers $(k_x,k_y,b_x,b_y,c)$, and their
dimensions $l_R$. The symbol $*$ refers to indefinite reflection
quantum numbers.
}
\end{table}

The irreducible representations $R$ of the space groups $G_L$ 
with $L=3,5,6$ can be derived analytically following
Ref.~\onlinecite{Fano}. In Table~\ref{tab:Irrep356} each of them
are specified by the lattice size $L$ and by their representation
quantum numbers $(k_x,k_y,b_x,b_y,c)$ related to the
translation and reflection operators $\{t_x,t_y,r_x,r_y,r_d\}$,
respectively. Furthermore, the dimensions $l_R$ of the representations
are listed.

\section{Projection operators of continuous symmetries}
\label{sec:appendixC}

There exists several methods for calculating the projection operators
corresponding to a continuous symmetry given by a Hermitian
operator $\hat{O}$; here we think of $\hat{O}$ being either the spin
operator $\hat{S}^2$ or the pseudospin operator $\hat{J}^2$. In this
Appendix we present an algorithm based on successive applications of
$\hat{O}$. Let $|\phi_0\rangle$ be a given normalized state we want to
project into $\hat{O}$-symmetry invariant subspaces. When $\hat{O}$
acts on $|\phi_0\rangle$ a term proportional with $|\phi_0\rangle$ is
generated together with a rest term. The rest term is denoted $f_1
|\phi_1\rangle$, and it defines a new unit vector $|\phi_1\rangle$
perpendicular to $|\phi_0\rangle$ while $f_1$ is a prefactor: 

\begin{equation}
\hat{O} |\phi_0\rangle = 
\langle \phi_0|\hat{O}|\phi_0\rangle |\phi_0\rangle + 
f_1 |\phi_1\rangle.
\end{equation}
If $f_1 = 0$ we are done, if not, we continue by applying $\hat{O}$ to
$|\phi_1\rangle$ and expand the result on $|\phi_0\rangle$,
$|\phi_1\rangle$. The rest term is now denoted $f_2 |\phi_2\rangle$,
where $|\phi_2\rangle$ is a unit vector perpendicular to both
$|\phi_0\rangle$ and $|\phi_2\rangle$ and $f_2$ is a prefactor:

\begin{equation}
\hat{O} |\phi_1\rangle = 
\langle \phi_0|\hat{O}|\phi_1\rangle |\phi_0\rangle + 
\langle \phi_1|\hat{O}|\phi_1\rangle |\phi_1\rangle + 
f_2 |\phi_2\rangle.
\end{equation}
Since we are working in a finite Hilbert space this process is
guarantied to yield a zero rest term after $M$ steps, i.e.\ $f_M = 0$.
The set $S_{\hat{O}}(|\phi_0\rangle) = \{ |\phi_0\rangle,|\phi_1\rangle,
\ldots, |\phi_{M-1}\rangle \}$ thus yields the smallest
$\hat{O}$-invariant subspace containing the starting vector
$|\phi_0\rangle$. The symmetry operator $\hat{O}$ is then diagonalized
within $S_{\hat{O}}(|\phi_0\rangle)$ yielding the eigenvalues
$\omega_k$ and eigenvectors $|\omega_k\rangle$:

\begin{equation}
\hat{O} |\omega_k\rangle = \omega_k |\omega_k\rangle \hspace{10mm}
\mbox{with} \hspace{10mm} |\omega_k\rangle = \sum_{i=0}^{M-1} c_{ki}
|\phi_i\rangle.
\end{equation}
The projection ${\cal P}_{\hat{O}}^{\omega_k}$ of $|\phi_0\rangle$ into
the $\hat{O}$-symmetry invariant subspace corresponding to the
eigenvalue $\omega_k$ is thus simply:

\begin{equation}
{\cal P}_{\hat{O}}^{\omega_k}|\phi_0\rangle = c_{k0}^* |\omega_k\rangle.
\end{equation}
Based on this equation we find the expressions for the projections in
spin space, Eqs.~(\ref{eq:PS0})-(\ref{eq:PS2}), and in pseudospin space,
Eqs.~(\ref{eq:PJ00})-(\ref{eq:PJ22}).

\section{Simultaneous eigenstates of $\hat{T}$ and $\hat{U}$}
\label{sec:appendixD}
In this appendix we present the analytical construction of 
simultaneous eigenstates of $\hat{T}$ and $\hat{U}$. The existence of
these states explains the remaining degeneracies in the symmetry
invariant subspaces, and the number of them (found numerically) are
listed in Table~\ref{tab:DegRemain} grouped after the
$\hat{U}$-eigenvalue $\gamma$ (=0,1,2 for four-electron systems) and
the spin $S$. We denote these states $\hat{T}/\hat{U}$ states or
$|\psi_{S}^{\gamma}\rangle$. A priori, eigenstates of $\hat{U}$ 
is most conveniently described in real space whereas eigenstates of
$\hat{T}$ are naturally given in momentum space. To require a state to
be a $\hat{T}/\hat{U}$ state imposes severe constraints. Below we
find many of these states analytically. Since in the following we will
be using both real space states and momentum space states, we will
reserve the letters $a$, $b$, $c$, and $d$ for sites in real space and
the letters $k$, $q$, $p$ and $r$ for momenta. 

We begin from below in Table~\ref{tab:DegRemain} by studying two-pair
states, i.e.\ $\gamma=2$. In a naive 
sense, such a state must be a superposition of extremely localized
states in real space or correspondingly of very out-spread states in
momentum space. This involves superpositions of many states with
different wave vectors $\bf k$ and hence different energies
given by Eq.~(\ref{eq:ZeroEigen}). It turns out that to 
obtain an energy eigenstate states only states where $\bf k$ enters
together with $\bfpi - {\bf k}$, where $\bfpi = (\pi,\pi)$, can be used
since $\cos(k^{\zeta}) + \cos(\pi-k^{\zeta}) = 0$, and consequently
these states only exist for even $L$. We construct the desired state
$|\psi^{\gamma=2}_{S=0}\rangle$ by superposing two-pair states:

\begin{equation} \label{eq:PsiG2S0}
|\psi^{\gamma=2}_{S=0} \rangle \equiv 
\sum_{ab} e^{i\bfspi \cdot {\bf a}} e^{i\bfspi \cdot {\bf b}} 
|a,b;a,b\rangle = 
\sum_{\bf kp} |{\bf k},{\bf p};(\bfpi-{\bf k}),(\bfpi-{\bf p}) \rangle.
\end{equation}
From the real space representation it is seen immediately that
$\hat{U}|\psi^{\gamma=2}_{S=0}\rangle=2|\psi^{\gamma=2}_{S=0}\rangle$
and that $\hat{\cal P}^{(0)}_S |\psi^{\gamma=2}_{S=0} \rangle =
|\psi^{\gamma=2}_{S=0} \rangle$, while the momentum space
representation directly yields 
$\hat{T}|\psi^{\gamma=2}_{S=0}\rangle=0|\psi^{\gamma=2}_{S=0}\rangle$.
In fact, as can be seen in the rows $\gamma=2$ of
Table~\ref{tab:DegRemain}, this is the only $\hat{T}/\hat{U}$ state
with $\gamma = 2$, e.g.\ it is easily seen from 
Eqs.~(\ref{eq:PS0})-(\ref{eq:PS2}) that no $\gamma$=2 state can have
$S$=1 or 2. Note how $|\psi^{\gamma=2}_{S=0} \rangle$ is
independent of the coupling strength $U$, but that its energy is
$U$-dependent and of the form $E(U) = E(0) + 2 U$. 

Having explained the $\gamma$=2 rows of Table~\ref{tab:DegRemain} we
turn to the $\gamma$=1 rows. From Eqs.~(\ref{eq:PS0})-(\ref{eq:PS2})
we find that there exists no 
$\gamma$=1 states with $S$=2. To find the $\gamma$=1 states with
$S$=0,1 we construct states $|\psi_{\bf kq}\rangle$ which manifestly
contains exactly one pair, such that $\hat{U} |\psi_{\bf kq}\rangle =
|\psi_{\bf kq}\rangle$: 

\begin{eqnarray}
|\psi_{\bf kq}\rangle & = & \sum_a e^{i\bfspi \cdot {\bf a}} 
e^{i[{\bf k}\cdot {\bf b} + ({\bf q}-{\bf k})\cdot {\bf d}]} 
(1 - \delta_{b,d}) | a,b;a,d \rangle \nonumber \\
\label{eq:psikq} &= & \sum_{\bf p} 
|{\bf p},{\bf k};(\bfpi-{\bf p}),({\bf q}-{\bf k})\rangle
- \frac{1}{L^2} \sum_{\bf pr}
|{\bf p},{\bf r};(\bfpi-{\bf p}),({\bf q}-{\bf r})\rangle.
\end{eqnarray}
Note how the double sum in Eq.~(\ref{eq:psikq}) involves all ${\bf k}$
vectors but that it only depends on {\bf q}. To obtain $S$=1 we simply
form anti-symmetric combinations of such states: 

\begin{eqnarray}
|\psi^-_{\bf kq}\rangle & = & 
|\psi_{\bf kq}\rangle - |\psi_{({\bf q}-{\bf k}){\bf q}}\rangle 
\nonumber \\
\label{eq:Psi-} & = & \sum_{\bf p}  (
|{\bf p},{\bf k};(\bfpi-{\bf p}),({\bf q}-{\bf k})\rangle -
|{\bf p},({\bf q}-{\bf k});(\bfpi-{\bf p}),{\bf k}\rangle).
\end{eqnarray}
Firstly, we note that $|\psi^-_{\bf kq}\rangle \neq 0$ if and only if
${\bf q} \neq 2{\bf k}$. Secondly, since only permutations of the
vectors ${\bf k}$ and $({\bf q} - {\bf k})$ are involved we find that 
$\hat{H} |\psi^-_{\bf kq}\rangle = E |\psi^-_{\bf kq}\rangle$. And thirdly, 
$\hat{\cal P}^{(1)}_S|\psi^-_{\bf kq}\rangle = - |\psi^-_{\bf kq}\rangle$.
Thus we have found $(\stackrel{L^2}{_2})$ $\hat{T}/\hat{U}$ states
$|\psi^{\gamma=1}_{S=1} \rangle$. This yields exactly the number of
states listed in the $\gamma$=1/$S$=1 row of
Table~\ref{tab:DegRemain}. 

\begin{equation} \label{eq:PsiG1S1}
|\psi^{\gamma=1}_{S=1} \rangle = |\psi^-_{\bf kq}\rangle.
\end{equation}
The states $|\psi^{\gamma=1}_{S=0} \rangle$ must be sought
among linear combinations of $|\psi^+_{\bf kq}\rangle$ defined as 

\begin{eqnarray} \label{eq:Psi+}
|\psi^+_{\bf kq} \rangle & = & 
|\psi_{\bf kq}\rangle + |\psi_{({\bf q}-{\bf k}){\bf q}}\rangle
\\
\nonumber
& = & {\displaystyle \sum_{\bf p} \left[
|{\bf p},{\bf k};(\bfpi\!-\!{\bf p}),({\bf q}\!-\!{\bf k})\rangle \!+\! 
|{\bf p},({\bf q}\!-\!{\bf k});(\bfpi\!-\!{\bf p}),{\bf k}\rangle\right]}
{\displaystyle - \frac{2}{L^2} \sum_{\bf pr}
|{\bf p},{\bf r};(\bfpi\!-\!{\bf p}),({\bf q}\!-\!{\bf r})\rangle.}
\end{eqnarray}
The problem now, however, is that the double sum does not vanish, hence
preventing the state of being an energy eigenstate. Only
upon forming differences $|\psi^+_{\bf kq}\rangle - |\psi^+_{\bf
k'q}\rangle$ with ${\bf k'} \neq {\bf k}$ can we get rid of it. But
the resulting state will only be an energy eigenstate if 
$\sum_{\zeta}[\cos(k^{\zeta}) + \cos(q^{\zeta}\!-\!k^{\zeta})]$ equals
$\sum_{\zeta}[\cos(k'^{\zeta}) + \cos(q^{\zeta}\!-\!k'^{\zeta})]$.
This is easily obtained if ${\bf k'} = (q^x\!-\!k^x,k^y)$, since then we
are only permuting the momentum components. 

\begin{equation} \label{eq:PsiG1S0}
|\psi^{\gamma=1}_{S=0} \rangle = 
|\psi^+_{\bf kq}\rangle - |\psi^+_{\bf k'q}\rangle.
\end{equation}
By accident there can also exist
other values of ${\bf k'}$ which fulfills the requirement, and besides
combine pairs such as $|\psi^+_{\bf kq}\rangle - |\psi^+_{\bf
k'q}\rangle$ we can also in some cases combine three states, 
$2|\psi^+_{\bf kq}\rangle-(|\psi^+_{\bf pq}\rangle + |\psi^+_{\bf
p'q}\rangle)$, or four, $(|\psi^+_{\bf kq}\rangle + |\psi^+_{\bf
k'q}\rangle) - (|\psi^+_{\bf pq}\rangle + |\psi^+_{\bf p'q}\rangle)$,
or even more. By a straightforward combinatorial search
we find the number of states listed in the $\gamma$=1/$S$=0 row of
Table~\ref{tab:DegRemain}, and we have thus identified all states in
the $\gamma$=1 rows, and found them to be independent of $U$ but with
an energy dependence of the form $E(U) = E(0) + U$.

Finally, we turn to the $\gamma$=0 rows of Table~\ref{tab:DegRemain}.
First we note that all $S$=2 states of the systems are found here.
This is easily proved by noting that the states with $(S,S_z) = (2,0)$
are formed by applying the spin lowering operator $S_-$ (which
commutes with $H$) twice to states with $(S,S_z) = (2,2)$. But the
latter states cannot contain any doubly occupied sites since that
would yield a lower than maximal value of $S_z$. Clearly, these
states as well as their energies are independent of $U$.
Then we consider a large class of energy eigenstates with
$S=0,1$ and $\gamma=0$, which does not require the momentum component $\pi$, 
and which therefore accounts for degeneracies for any value of $L$.
Writing $|(k^x,k^y)\rangle = |k^x\rangle \!\otimes\! |k^y\rangle$
we construct $\gamma$=0 states $|\phi\rangle$ obeying
$\hat{U}|\phi\rangle=0$ by symmetrizing one component, say the $x$
component, and letting $\hat{\cal P}^{(2)}_S$ act on the other:

\begin{eqnarray} \label{eq:phi}
{\displaystyle
|\phi\rangle} & = &
{\displaystyle
|k^x_1,k^x_2;k^x_3,k^x_4\rangle_{\rm sym}
\otimes \hat{\cal P}^{(2)}_S|k^y_1,k^y_2;k^y_3,k^y_4 \rangle,}
\qquad {\rm with}\\
\nonumber
{\displaystyle |k^x_1,k^x_2;k^x_3,k^x_4\rangle_{\rm sym}} 
& \equiv &
{\displaystyle
|k^x_1,k^x_2;k^x_3,k^x_4\rangle+
|k^x_1,k^x_2;k^x_4,k^x_3\rangle+
|k^x_2,k^x_1;k^x_3,k^x_4\rangle+
|k^x_2,k^x_1;k^x_4,k^x_3\rangle}.
\end{eqnarray}
Since only momentum permutations enter, $|\phi\rangle$ is clearly an
energy eigenstate. The proper spin states are found by the standard
projections:

\begin{equation} \label{eq:PsiG0SS}
|\psi_{S=S'}^{\gamma=0} \rangle = \hat{\cal P}^{(S')}_S |\phi\rangle.
\end{equation}
Simple combinatorics yields 0,30,300 and 1680 $S$=0 states and
0,90,1050, and 6300 $S$=1 states for $L = 3,4,5$ and 6 respectively.
In analogy with the $\gamma$=1 case many more $\hat{T}/\hat{U}$ states
can be constructed for $L$=4 and 6 and $\gamma$=0 when the momentum
vector $\pi$ is taken into account. We give one example of a class of
such states. For a given momentum vector ${\bf k} = (k^x,k^y)$
we define for $\delta=x,y,xy$ the functions $\pi_\delta({\bf k})$,

\begin{equation} \label{eq:pidelta}
\pi_x({\bf    k}) = ( k^x\!+\!\pi, k^y        ), \;\;
\pi_y({\bf    k}) = ( k^x        , k^y\!+\!\pi), \;\;
\pi_{xy}({\bf k}) = ( k^x\!+\!\pi, k^y\!+\!\pi),
\end{equation}
based on which we introduce
two operators $\hat{\Pi}^{\sigma,\sigma}_\delta$ and 
$\hat{\Pi}^{\bar{\sigma},\sigma}_\delta$:

\begin{equation} \label{eq:PiSigma}
\begin{array}{rclcl}
\hat{\Pi}^{ \sigma,\sigma}_\delta
|{\bf k}_1,{\bf k}_2;{\bf k}_3,{\bf k}_4\rangle = & + &
|{\bf k}_1,{\bf k}_2;{\bf k}_3,{\bf k}_4\rangle & + &
|\pi_\delta({\bf k}_1),\pi_\delta({\bf k}_2);{\bf k}_3,{\bf k}_4\rangle
\\ & + & 
|{\bf k}_1,{\bf k}_2;\pi_\delta({\bf k}_3),\pi_\delta({\bf k}_4)\rangle &+&
|\pi_\delta({\bf k}_1),\pi_\delta({\bf k}_2);
 \pi_\delta({\bf k}_3),\pi_\delta({\bf k}_4)\rangle,\\[2mm]
\hat{\Pi}^{\bar{\sigma},\sigma}_\delta
|{\bf k}_1,{\bf k}_2;{\bf k}_3,{\bf k}_4\rangle = & + &
|\pi_\delta({\bf k}_1),{\bf k}_2;\pi_\delta({\bf k}_3),{\bf k}_4\rangle &+&
|\pi_\delta({\bf k}_1),{\bf k}_2;{\bf k}_3,\pi_\delta({\bf k}_4)\rangle
\\ &+&
|{\bf k}_1,\pi_\delta({\bf k}_2);\pi_\delta({\bf k}_3),{\bf k}_4\rangle &+&
|{\bf k}_1,\pi_\delta({\bf k}_2);{\bf k}_3,\pi_\delta({\bf k}_4)\rangle.\\
\end{array}
\end{equation}
Direct inspection shows $\hat{U}
(\hat{\Pi}^{\sigma,\sigma}_\delta -
\hat{\Pi}^{\bar{\sigma},\sigma}_\delta) 
|{\bf k}_1,{\bf k}_2;{\bf k}_3,{\bf k}_4\rangle = 0$, and by enforcing 
certain constraints on all eight momentum components this state also
becomes an energy eigenstate with energy $E$=0, while applying the
projector $\hat{\cal P}_S^{(S')}$ renders the correct spin $S$=$S'$:

\begin{equation} \label{eq:PsiG0SSpi}
|\psi_{S=S'}^{\gamma=0} \rangle = \hat{\cal P}^{(S')}_S 
(\hat{\Pi}^{\sigma,\sigma}_\delta - \hat{\Pi}^{\bar{\sigma},\sigma}_\delta)
|{\bf k}_1,{\bf k}_2;{\bf k}_3,{\bf k}_4\rangle, \;
{\rm with} \; k^x_n = \pi\!-\!k^y_n, \; n=1,2,3,4.
\end{equation}
Finally, we note that for lattices containing the momentum $\pi/2$ as
is the case for $L=4$, even more $\hat{T}/\hat{U}$ states can be
constructed, in accordance with Table~\ref{tab:DegRemain}. An example
of this can be obtained from Eq.~(\ref{eq:PsiG0SSpi}). If for example
we let $\delta=x$ then it suffices to enforce the constraint $k^x_n =
\pm\pi/2$ while allowing any value for the $y$ components. The result
is energy eigenstates with energy $E=\sum_n \cos(k^y_n)$.

We conclude that many of the $\hat{T}/\hat{U}$ states 
$|\psi^{\gamma}_{S}\rangle$ found numerically have been constructed
analytically, and in agreement with the numerical findings all these
states are independent of $U$, while their energies are of the form
$E(U) = E(0) + \gamma U$. The analytic constructions reveal that these
states are due to a restricted permutation symmetry for the momentum
components of states in momentum space.


\begin{thebibliography}{99}
\bibitem{direct} A group $G$ is a direct product group if it possesses
two subgroups $G_1$ and $G_2$ such that ($i$) $\forall g \!\in\! G
\;\exists g_1 \!\in\! G_1 \;\exists g_2 \!\in\! G_2\!: g = g_1 g_2$, ($ii$)
$G_1 \cap G_2 = {e}$ ($e$ is the identity), ($iii$) $\forall
g_1 \!\in\! G_1 \;\forall g_2 \!\in\! G_2\!: g_1 g_2 = g_2 g_1$. One then
writes $G = G_1 \otimes G_2$.
\bibitem{semidir} A group $G$ is a semi-direct product group if it
possesses two subgroups $G_1$ and $G_2$ such that ($i$) $\forall 
g \!\in\! G \;\exists g_1 \!\in\! G_1 \;\exists g_2 \!\in\! G_2\!: g = g_1
g_2$, ($ii$) $G_1 \cap G_2 = {e}$, ($iii$) $G_1$ is an invariant
subgroup, i.e.\ $\forall g_1 \!\in\! G_1 \;\forall g \!\in\! G\!: g
g_1 g^{-1} \!\in\! G_1$. One then writes $G = G_1 \semidir G_2$.
\bibitem{TwoGen} Using standard cycle-notation an example of two
permutations generating all of $G_4$ is (4,12)(5,13)(6,14)(7,15) and
(0,4,8,9,10,14,2,3)(1,7,12,5,11,13,6,15).
\bibitem{SymHubModel} The symmetrized Hubbard model $\hat{H}'$ is
obtained by the substitution \\
$\hat{U} \rightarrow \hat{U}' = \sum_i (\hat{n}_{i\uparrow} - \frac{1}{2})
(\hat{n}_{i\downarrow} - \frac{1}{2})$ leading to $\hat{H}' = 
\hat{H} + u(L^2/4 - \sum_{i\sigma} \hat{n}_{i\sigma}/2)$.
\bibitem{SvsJ} The pseudospin symmetry can be viewed as a consequence
of a particle-hole symmetry which maps a Hubbard model $H$ with
coupling strength $U$ and $n_u$ ($n_d$) spin-up (spin-down) electrons
onto a Hubbard model $\tilde{H}$ with $\tilde{U} = -U$, $\tilde{n}_u =
n_u$, and $\tilde{n}_d = L^2-n_d$. The pseudospin of $H$ is the spin
of $\tilde{H}$ and vice versa \cite{Lieb}.
\end{thebibliography}
\end{document}